\documentclass[]{aastex631}

\usepackage{bm}
\usepackage{subfigure}
\usepackage{amsmath}
\usepackage{cases}
\usepackage{comment}
\usepackage{soul}

\DeclareUnicodeCharacter{2212}{-}
\usepackage{newunicodechar}

\newunicodechar{≤}{\ensuremath{\leq}}

\usepackage{xcolor}

\newcommand{\Bell}{\ensuremath{\boldsymbol\ell}}
\newcommand{\DS}{\displaystyle}

\shorttitle{SNR interaction with a turbulent interstellar background}
\shortauthors{Prete et al.}
\graphicspath{{./}{figures/}}

\begin{document}

\nolinenumbers
\title{The Interaction of a Supernova Remnant with background interstellar turbulence}

\author[0000-0003-3739-3170]{Giuseppe Prete}
\affiliation{Università della Calabria via P. Bucci, cubo 33C, 87036, Rende (CS), Italy}

\author{Silvia Perri}
\affiliation{Università della Calabria via P. Bucci, cubo 33C, 87036, Rende (CS), Italy}
\affiliation{National Institute for Astrophysics (INAF), Scientific Directorate, via del Parco Mellini 84, Roma, 00136, RM, Italy.\textbf{}}

\author{Claudio Meringolo}
\affiliation{Institut für Theoretische Physik, Goethe Universität, Max-von-Laue-Str. 1, D-60438 Frankfurt am Main, DE}

\author{Leonardo Primavera}
\affiliation{Università della Calabria via P. Bucci, cubo 33C, 87036, Rende (CS), Italy}
\affiliation{National Institute for Astrophysics (INAF), Scientific Directorate, via del Parco Mellini 84, Roma, 00136, RM, Italy.\textbf{}}

\author{Sergio Servidio}
\affiliation{Università della Calabria via P. Bucci, cubo 33C, 87036, Rende (CS), Italy}
\affiliation{National Institute for Astrophysics (INAF), Scientific Directorate, via del Parco Mellini 84, Roma, 00136, RM, Italy.\textbf{}}



\begin{abstract}
Supernova explosions (SNe) are among the most energetic events in the Universe. 
After the explosion, the material ejected by the Supernova expands throughout the interstellar medium (ISM) forming what is called Supernova Remnant (SNR). Shocks associated with the expanding SNR are sources of galactic cosmic rays, that can reach energy of the PeV order. In these processes, a key role is played by the magnetic field. It is known that the ISM is turbulent with an observed magnetic field of about a few $\mu$G, made by the superposition of a uniform and a fluctuating component. During the SNR expansion, the shock interacts with a turbulent environment, leading to a distortion of the shock front and a compression of the medium.
In this work, we use the MagnetoHydroDynamics (MHD) PLUTO code to mimic the evolution of the blast wave associated with the SNR. We make a parametric study varying the level of density and magnetic field fluctuations in the interstellar medium, with the aim of understanding the best parameter values able to reproduce real observations. We introduce a novel analysis technique based on two-dimensional autocorrelation function $C{(\bf{\Bell})}$ and the second order structure function $S_2{(\bf{\Bell})}$, quantifying the level of anisotropy and the turbulence correlation lengths. By interpolating the autocorrelation function on a polar grid, we extract the power spectra of turbulence at the SNR.
Finally, a preliminary comparison with Chandra observations of SN 1006 is also presented.

\end{abstract}


\section{Introduction}
A Supernovae explosion (SNe) is an instantaneous release of energy of about $\simeq 10^{51} $ erg, associated to the catastrophic collapse of a massive star or due to a  runaway nuclear burning of a white dwarf. 
 
After SNe, the material ejected from its progenitor permeates the interstellar medium (ISM), expanding at a speed that can exceed the local sound speed. This gives rise to a shock front, also known as a forward shock (FS), which tends to be slowed down by its interaction with the ISM. Thus, the FS sweeps up material and gets decelerated. The ejecta tends to catch the shock and a reverse shock (RS) is generated, propagating backward in the reference frame of the ejecta. Between these two shocks an interface region forms, called Contact Discontinuity (CD). It is a boundary region between a part of the ejecta heated by the RS and the other part heated by the FS-ISM interaction. The expanding material collected by the blast wave during its evolution through the ISM gives rise to the Supernova Remnant (SNR). Three phases can be recognized in the SNR evolution:  1) the free expansion phase, characterized by an FS moving at almost constant speed into the ISM (in this phase the amount of material swept up by the shock is smaller than the mass of the ejecta); 2) the Sedov-Taylor phase (or energy-conserving phase) when the mass of the swept-up gas exceeds the mass of the ejecta and the kinetic energy of the explosion is transferred to the swept up material, which does not radiate; 3) the radiative phase dominated by the radiation losses, involving several emission lines, during which the shock slows down and the gas becomes cooler \citep{Raymond-1979,Helder_2012}.

It is widely believed that SNR shocks are the sources of Galactic cosmic rays, whose energy can reach $10^{15}$ eV, and also sources of X-ray and radio emissions. These high Mach number shocks can accelerate energetic particles at relativistic energies through the diffusive shock acceleration (DSA) mechanism, in which particles can cross the shock back and forth due to the scattering with magnetic irregularities, gaining an amount of energy at each crossing (\citealp{fermi1949origin, Krymskii1977, Blandford1978, Bell-1978a, Bell-1978b, Drury1983,jokipii1982particle, jokipii1987rate,guo2010particle, liu2021acceleration}). In these high-energy processes, a key role is played by the magnetic field. It is known that the ISM is turbulent with a Kolmogorov-like power spectrum (\citealp{Lee1976, Armstrong1981, Armstrong1995, chepurnov2010extending, linsky2022inhomogeneity-a, linsky2022inhomogeneity-b}). The observations have revealed that the Galactic magnetic field assumes values of few $\mu$G and is given by the superposition of a uniform and a random/fluctuating component, which are approximately in equipartition \citep{Beck1996, Minter1996, Han2004}. Turbulence is one of the most important physical processes in astrophysics, spanning from cosmology to the interstellar medium, including stars, supernovae, accretion disk, and many others \citep{radice2018turbulence, meringolo2023microphysical, wang2024role}. It is ubiquitous in the Universe and the dynamics of non-stationary astrophysical plasma include a wide range of length scales, velocities, and timescales. Hence, the study of this kind of phenomenon can have a key role in the generation of large-scale magnetic fields that characterize many celestial objects. The ISM turbulence scales range from kpc to AUs (\citet{Armstrong1995, Elmegreen_2004,Brandenburg_2013}). MHD turbulence therefore plays a fundamental role in astrophysical processes in the ISM. The role of MHD turbulence has been tested during the last decades with the support of numerical simulations which confirmed many theoretical predictions (\citealp{kolmogorov1941,shebalin1983anisotropy, Cho_2005,kritsuk2007statistics, federrath2010comparing, federrath2016universality, ferrand2019exact}).

Several studies have investigated the interaction between turbulent magnetic fields and shock waves.
Such an interaction can distort the shock surface, leading to the formation of shock ripples \citep{ofman2013rippled,Johlander16}, and to an increase of the fluctuation level in the downstream region \citep{Neugebauer2005,lu2009interaction,trotta2021phase,Nakanotani22}. 
The process of magnetic field amplification due to the turbulent dynamo has widely been explored, making use of MHD simulations with 1D \citep{Dickel1989}, 2D \citep{Giacalone_2007,inoue2009turbulence,guo2012amplification,fraschetti2013turbulent, mizuno2014magnetic} and 3D \citep{inoue2013origin,ji2016efficiency,hu2022turbulent} cofigurations.  These works show that turbulence can amplify the magnetic field in the postshock region by a factor of 100.

The interaction between shocks associated with SNR and the turbulent ISM has been deeply studied in (\citet{guo2012amplification,yu2015three, velazquez2017}). In particular, \citet{guo2012amplification} developed a 2D MHD simulation in which they studied the amplification of the magnetic field when a supernova blast wave propagates into a turbulent ISM. Two different processes can affect the magnetic field amplification: 1) the interaction between the shock front and the turbulent environment that allows the formation of the shock ripples, which amplify the magnetic field; 2) the growth of the Rayleigh-Taylor Instability (RTI) at the contact discontinuity region, that produces the highest amplification of the magnetic field. 
RTI occurs when two fluids of different densities interact. When an impulsively accelerated shock wave propagates between two fluids with different densities, an evolution of the RTI appears, called Richtmyer–Meshkov instability (RMI) \citep{richtmyer1954taylor,meshkov1969instability}.
RMI is believed to be responsible for the amplification of the magnetic field in SNRs \citep{sano2012magnetic,nagel2017platform}. 
A secondary instability, namely the Kelvin–Helmholtz instability (KHI), can also induce turbulence. It develops when a shear of velocity appears at the interface between two fluids \citep{helmholtz1868,thomson1871}. KHI plays an important role in various regions of the solar system, such as planetary magnetopauses, solar atmosphere, solar wind but also in astrophysical systems such as pulsar wind nebulae and around quasar \citep{masson2018}.

\citet{hu2022turbulent} compared 2D and 3D MHD simulations of shock propagation inside a dense turbulent environment. They find differences in terms of morphology and amplification of the magnetic field. 3D simulations show larger amplification than 2D and, as a consequence, the morphology of the shock looks quite different, with a more corrugated shock front and the development of filaments. 
\\
To better understand the influence of turbulence on the transport properties of the blast wave in SNR, it is fundamental to explore all the possible configurations of the ISM background, making a parametric study of the shock-turbulence interaction, in order to ``tune'' the simulation parameters for comparison with observations.

Particle acceleration in SNR shocks is testified by the presence of synchrotron emission from relativistic electrons. The emission is brighter at the edges of the blast waves, giving rise to bright rims \citep{Morlino2010}, where particle acceleration processes occur. The magnetic field can also be amplified by the presence of energetic particles streaming back ahead of the shock \citep{amato2014origin}  exciting low-frequency waves that enhance locally the magnetic field \citep{bell2004turbulent}. Actually, the presence of turbulence in the upstream region can be the cause of the irregular emissions seen in SNRs and also of the particular shape that the edges of the shock assume when they interact with the ISM \citep{Anderson1996,balsara2001evolution,guo2012amplification,reynoso2013radio,yu2015three}.
Such an irregular and non-uniform emission has been observed in SN 1006. It is a Type Ia supernova, i.e. the result of the explosion of a white dwarf in an accretion binary system. In these systems, white dwarfs increase their mass from a companion that can be any type of star. The companion pushes the white dwarf over the Chandrasekhar limit into core collapse and this results in a total disruption of the star. There is no stellar remnant in this case, such as a neutron star or black hole \citep{weiler1988supernovae}. SN 1006 has a bilateral morphology consisting of two bright limbs located in the north-eastern part (NE) and in the south-western part (SW), that present knots and filaments along the boundary of the remnant \citep{koyama1995evidence, bamba2003small,rothenflug2004geometry,reynoso2013radio}. The NE and the SW regions mainly emit in the radio and X-ray bands, which correspond to the synchrotron emission from high-energy electrons, while the $\gamma$-ray band could be associated with high-energy protons. Synchrotron emission is responsible for both the non-thermal emission in the radio band and the X-ray band.

This paper aims to mimic the interaction between a shock associated with the expansion of a SNR within a turbulent ISM. Since the turbulence properties of the ISM around a SNR shock are difficult to infer, we will make use of numerical simulations to understand the turbulent conditions of the ISM that better reproduce observations. In particular, we focus on SN 1006, which is an old SNR largely studied in literature (\citet{bamba2003small, cassam2008morphological, miceli2009thermal, orlando2012role,perri2016transport,  orlando2021modeling}). We aim to understand how the turbulent ISM influences the SN 1006 expansion by systematically varying the properties of turbulence.   

The evolution of SN 1006 is reproduced by using the MHD PLUTO code \citep{mignone2007pluto, mignone2011pluto}, setting different turbulent conditions in the ISM, both in the density and in the magnetic fields.

We perform a parametric study that allows us to determine the factors that mostly affect the SNR expansion.
Finally, we implement an analysis of the turbulence anisotropy within the remnant, after its interaction with the ISM, by means of a novel technique to estimate the power spectra. It is based on the study of the 2D autocorrelation function. We apply a ring-shaped mask to the SNR and we extract the 2D autocorrelation function and the structure function. We then interpolate both into polar coordinates in order to measure the correlation length at fixed angle values. For each angle, we compute the power spectra by applying the Blackman-Tuckey technique \citep{blackman1958measurement,blackman1958measurement2}.

Such a procedure will also be applied to {\it Chandra} observations of SN 1006 in the energy band (1-3 keV) for qualitative comparison with numerical simulations.


\section{Numerical model}

To reproduce the dynamical evolution of a SNR blast wave propagating through a magnetized environment we use the time-dependent, ideal, MHD equations of mass, momentum, and energy conservation in a Cartesian coordinate system. We choose to integrate these equations by using the PLUTO code \citep{mignone2007pluto, mignone2011pluto}. PLUTO is a Godunov-type code that provides a modular computational framework for the solution of gas dynamics equations under different regimes. 



The governing MHD equations used in the code are 

	
		
		

\begin{equation}
\begin{aligned}
	\frac{\DS\partial \rho}{\DS\partial t} + \bm{\nabla} \cdot (\rho \bm{v}) = 0, \\
	\frac{\DS\partial (\rho \bm{v})}{\DS\partial t} + \bm{\nabla} \cdot ( \rho \bm{vv} - \bm{BB} ) + \bm{\nabla} P_t  = 0 ,\\
	\frac{\DS\partial \varepsilon}{\DS\partial t} +  \bm{\nabla} \cdot \biggl[ (\varepsilon + P_t) \bm{v} - (\bm{v} \cdot \bm{B} ) \bm{B} \biggr] = 0,\\
	\frac{\DS\partial \bm{B}}{\DS\partial t} - \bm{\nabla}  \times(\bm{ v} \times \bm{B} ) = 0.
\end{aligned}
\label{eqmhd}
\end{equation}

Equations (\ref{eqmhd}) are dimensionless, normalized to typical length ($L_0$= 1pc), velocity ($v_0$= $10^5$ cm/s) and density ($\rho_0$= 1.67 $\time 10^{24}$ gr/cm$^3$), as usually done in the PLUTO code \citep{mignone2007pluto}.
Here $\rho = \mu n_{tot} m_p$ is the mass density, where $\mu$ is the mean molecular weight, $n_{tot}$ the total number density of particle and $m_p$ the proton mass, $\bm{v}=(v_x,v_y,v_z)$  the fluid velocity, $\bm{B}=(B_x, B_y, B_z)$ the magnetic field, $P_t$ is the total gas pressure, i.e. the sum of the thermal and the magnetic pressure ($P_t = P + B^2/2$). 
At this stage, we do not take into account the effects related to thermal conduction, viscosity, and resistivity, radiative cooling, non-linear effects due to the acceleration of charged particles. We couple this set of equations with an equation of state for an ideal gas. The total energy density is $\varepsilon = \frac{P}{\gamma -1} + \frac{1}{2} \rho v^2 + \frac{B^2}{2}$, where $\gamma$ is the adiabatic index. The solutions of the MHD equations must satisfy the divergence-free condition, i.e. $\bm{\nabla} \cdot \bm{B} = 0$. This condition is not naturally preserved in numerical schemes without a proper technique. PLUTO uses three different techniques in order to fulfill the condition $\bm{\nabla} \cdot \bm{B} = 0$: the eight wave formulation \citep{powell1994approximate,Powell1999}, the hyperbolic divergence cleaning \citep{mignone2010second, mignone2010high} and the constrained transport (CT) \citep{balsara1999staggered,Londrillo2004}. In the eight waves formalism the magnetic field has a cell-centered representation and additional terms (source terms) are added to the ideal MHD equations and, depending on the Riemann solver used, the source terms can be discretized in different way. In the hyperbolic divergence cleaning the divergence-free condition is enforced by solving a modified system of conservation laws, in which the induction equation is coupled with a generalized Lagrange multiplier. This technique is useful because no source terms are introduced, so the equations keep a conservative form, all variables retain a cell-centered representation and it is possible to use different Riemann solvers. The CT formalism uses two sets of magnetic fields: a face-centered and a cell-centered. In the first one, the field components are located at different spatial points in the control volume while in the second one, the staggered magnetic field is treated as an area-weighted average on the zone face and Stokes' theorem is used to update it. In this work we advance the magnetic field and, to preserve the divergence-free condition, we use the CT formalism, though results are consistent when using the other two schemes.

\subsection{Numerical setup} \label{sec:numerical-setup}  
We set up simulations in two-dimensional Cartesian coordinates (x,y) with uniform grids. The computational domain extends from -30 pc to 30 pc in both directions sampled with a total of 4096$^2$ grid points. We set all the other quantities following \citet{orlando2012role} and \citet{guo2012amplification}. In particular, 
we assume an initial cylindrical remnant of radius $R_{SN,0} = 0.4$ pc and length in the z-direction of $L_z = 0.8$ pc, that corresponds to an age of $t_0 =$ 10 years, whose progenitor is a star with mass of 1.4 $M_{\odot}$. We set the internal energy within the cylinder area equal to $E_{inj}=1.5 \times 10^{51}$ erg. This allows us to define an internal initial number density of about $ n_{inj}= 13.3$ cm$^{-3}$. The environment in which our remnant is evolving is set with a uniform density $n = 0.1$ cm$^{-3}$ \citep{Morlino2010, perri2016transport} and a uniform temperature $T=10^4 $ K. The shock associated with the SNR is supposed to expand with a velocity $V_{sh} \simeq 4400$ km/s and the adiabatic index is chosen as $\gamma = 5/3$ for a monoatomic, non-relativistic, ideal gas. We follow the SNR expansion until 10 pc; at this stage, the supernova is in the Sedov-Taylor phase and the energy is conserved because we do not take into account radiative losses.
We set a background mean magnetic field $B_0 = 3 \mu $G, and we can vary its direction to span different case studies. The direction of the mean magnetic field in spherical coordinates is:
\[
\begin{cases}
B_{0x} = B_0 cos(\tilde{\phi}) sin(\tilde{\theta}) \\
B_{0y} = B_0 sin(\phi) sin(\tilde{\theta}) \\
B_{0z} = B_0 cos(\tilde{\theta}).
\end{cases}
\]

\noindent
In order to make a comparison with SN 1006 we choose $\phi = 150^{\circ}$ and $\tilde{\theta}=90^{\circ}$, where $\phi$ is the azimuthal angle and $\tilde{\theta}$ the co-latitude. These values are deduced from observations in \citet{reynoso2013radio}, where the mean magnetic field has been found to be oriented at 60$^{\circ}$ with respect to the Galactic plane normal. 


\subsection{Homogeneous turbulence}
\begin{figure}[!htbp]
\centering
\includegraphics[width=1\textwidth]{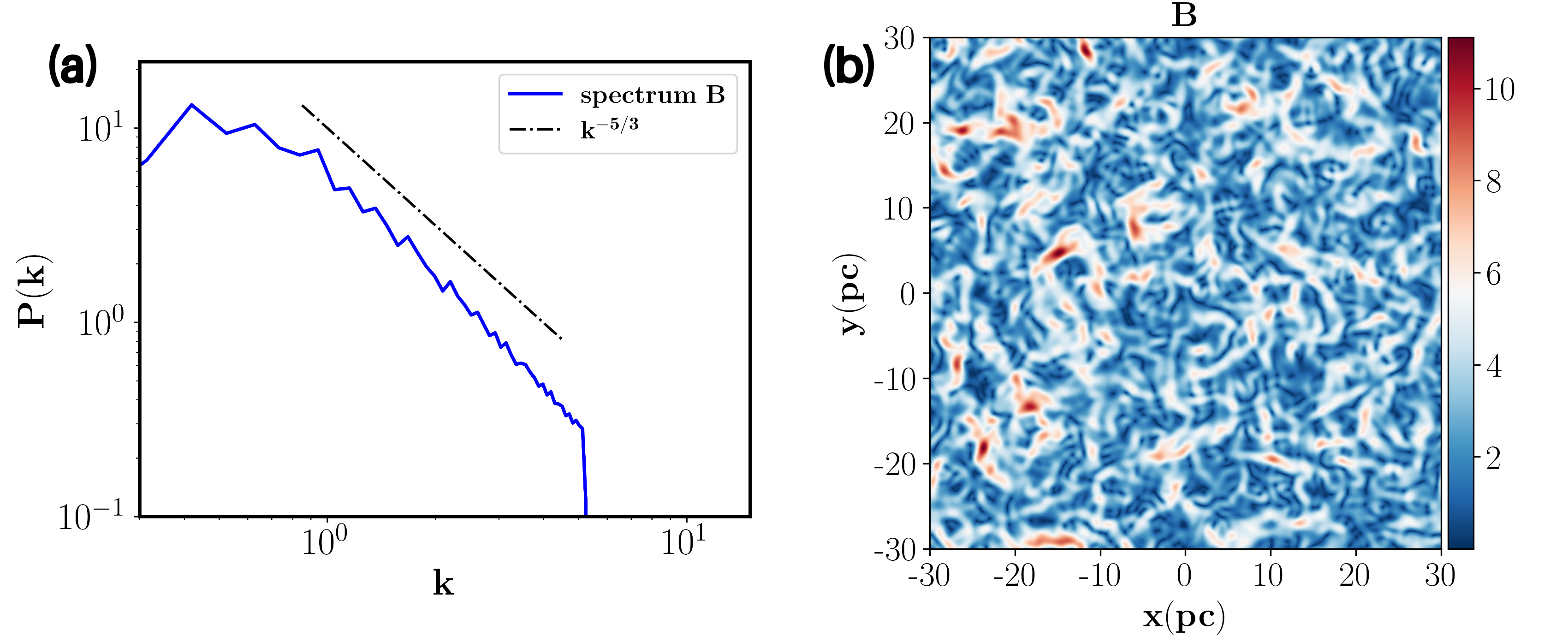}
\caption{a) Power spectral density of the total magnetic field (blue solid line). The dashed-dotted line represents the Kolmogorov scaling. (b) Initial magnetic field obtained by computing the curl of the $A_z$ potential vector.}
\label{fig:initial-cond}
\end{figure}

A compressible turbulent background in which the SNR blast wave propagates has been set as an initial condition of our simulations. Indeed, we consider the interstellar medium as made of density and magnetic field fluctuations superposed to a background field.  
We initialize the in-plane magnetic field as done in classical simulations of plasma turbulence \citep{servidio2012local,Meringolo2024}. Since we want to reproduce realistic conditions for the interstellar background medium, we consider different spectra for both fields. We start by expressing the $z-$component of the vector potential in Fourier modes as

\begin{equation}
    A_z(x,y) = \sum_{k_x, k_y} \widetilde{A}(k) e^{i( \bm{k} \cdot \bm{ x} + \phi_k)  } ,
    \label{Az}
\end{equation}

where $\bm{k}=(k_x, k_y)$ is the $k-$vector, $k=|\bm{k}|$, and $\tilde{\phi_k}$ are random phases.
In the above equation,  $\widetilde{A}(k) = [1+(k/k_0)^{\alpha}]^{-1}$ sets the initial shape of the spectrum, peaked at 
$k=k_0=(2 \pi/L)m_0$ with $m_0 = 4, L=60$ pc, and with a power-law $\alpha=15/6$.
The magnetic field is therefore computed as $\mathbf{B}=\nabla \times \mathbf{A}$, so that, according to equation (\ref{Az}), its power spectral density follows a Kolmogorov-like decay with $P(k)\sim k^{-5/3}$, as observed at intermediate scales in astrophysical plasma turbulence \citep{Bruno13}. Such an initial condition consists of random fluctuations on a wide range of scales.
We set a smooth cut-off in the Fourier space at $k^*=50$ to ensure that we can avoid possible aliasing effects at any used resolution \citep{meringolo2021spectral}. Figure \ref{fig:initial-cond}(a) shows the initial power spectrum of the total magnetic field. Note that the Kolmogorov spectrum is also plotted for comparison. In Figure \ref{fig:initial-cond}(b) we report the contours of the initial magnetic field, obtained from the $A_z$ potential vector, in the (x,y) plane.

\begin{figure}[!htbp]
\centering
\includegraphics[width=1\textwidth]{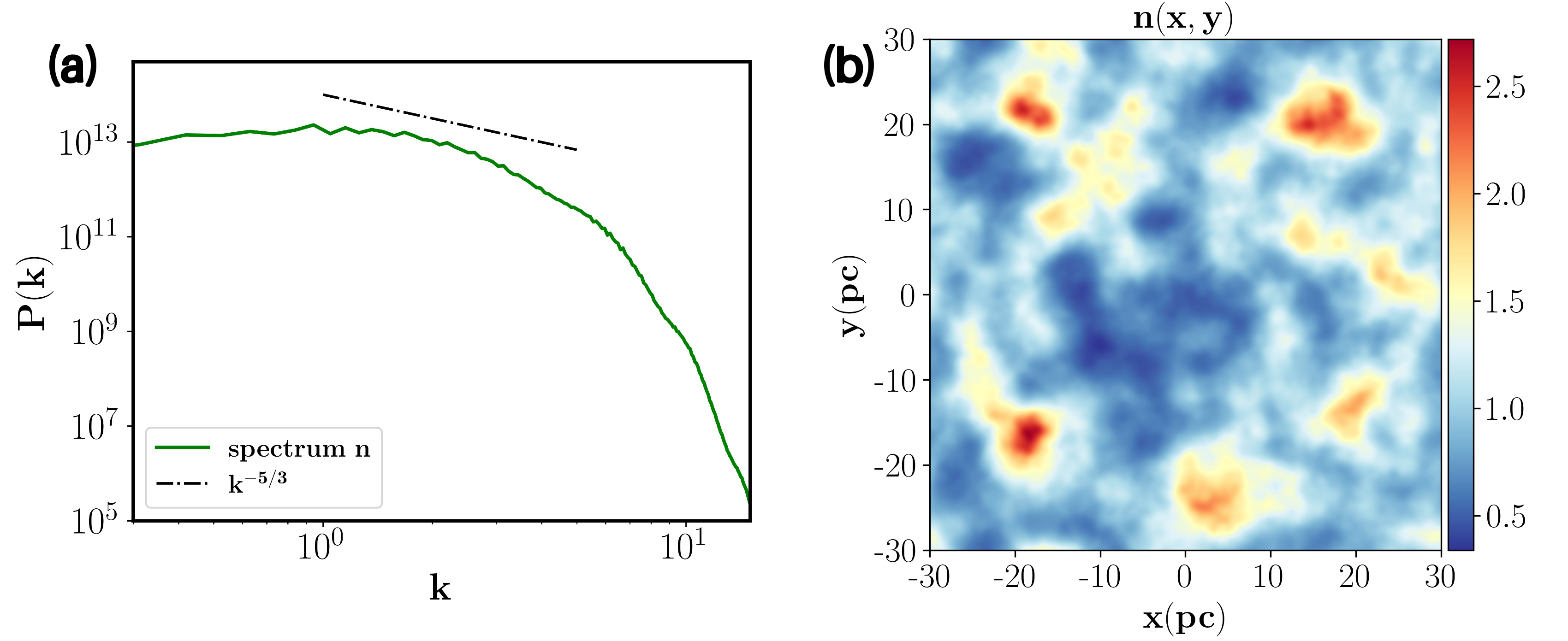}
\caption{(a) Power spectrum of the density (green solid line). (b) Background density map.}
\label{fig:initial-cond-rho}
\end{figure}

Since we know that in stellar wind, and in  particular in the solar wind, density fluctuations can assume a log-normal probability distribution \citep{burlaga2006magnetic}, we set the density field as following a log-normal distribution \citep{Giacalone_2007}
\begin{equation}
    n(x,y) = n_0 \exp(f_0+\delta f) ,
    \label{dens-eq}
\end{equation}

where $f_0$ is a properly chosen constant, and $\delta f$ is the density perturbation. We build the spectrum of $\delta f$ in a way analogous to the one we used for the vector potential in Eq.(\ref{Az}). Before applying the exponential function to the density fluctuations, we normalize the argument to unity in order to avoid extreme values. The spectrum of density fluctuations is plotted in Figure \ref{fig:initial-cond-rho}(a). The initial density field is also reported in Figure \ref{fig:initial-cond-rho}(b).

\section{Simulation Results} \label{sec:numerical results}
All the runs performed are summarized in Table \ref{table:table1}. In particular, we chose Run 2 (in bold in Table \ref{table:table1}) as a reference simulation. In run 2 the level of magnetic field turbulence is set to $\delta B/B=0.5$ \citep{guo2012amplification}, while the level of density fluctuations is $\delta n/n=1$. We fix the number density of the background close to the typical value of the ISM \citep{Morlino2010,perri2016transport}. 

\begin{table}[!htbp]
\centering
\begin{tabular}{c c c c c c c c c }
\hline
Run & $\bf{M_{ej}(M_{\odot})}$ & $\bf{n_{out}(cm^{-3})}$ &$  \text{\boldmath$\delta$} \bf{B/B}$ & $ \text{\boldmath$\delta$} \bf{n/n}$ &  $\bf{B_{avg} ( \text{\boldmath$\mu$} G)}$ & $\bf{B_{max} ( \text{\boldmath$\mu$} G)}$ & $\bf{n_{avg} (cm^{-3})}$ & $\bf{n_{max}(cm^{-3})}$ \\ 
\hline
1 & 1.4 & 0.1 & 0.1 & 1 & 4.88 & 425.1 & 0.42 & 2.29 \\
\bf{2 }& \bf{1.4} & \bf{0.1}& \bf{0.5} & \bf{1} & \bf{5.1} & \bf{475.5} & \bf{0.42} & \bf{2.29}  \\
3 & 1.4 & 0.1 & 1 &  1.0 & 5.8 & 521.6 & 0.42 & 2.29  \\
4 & 1.4 & 0.1 & 0.5 & 0.1 & 6.2 & 609.5 & 0.15 & 0.7 \\
5 & 1.4 & 0.1 & 0.5 & 0.5 & 5.5 & 474.2 & 0.27 & 1.43  \\
6 & 3 & 0.1 & 0.5 &  1 & 4.3 & 480.6 & 0.46 & 3.8 \\
7 & 5 & 0.1 & 0.5 & 1 & 3.82 & 415.9 & 0.5 & 6.99  \\
8 & 1.4 & 0.05 & 0.5 & 1 & 5.95 & 563.3 & 0.22 & 1.57  \\
9 & 1.4 & 1 & 0.5 & 1 & 3.4 & 74.7 & 4.03 & 18.26  \\
10 & 1.4 & 3 & 0.5 & 1 & 3.24 & 18.3 & 11.8 & 54.6  \\
\hline
\end{tabular}
\caption{List of the numerical runs performed. The values of the mass of the ejecta, the magnetic and the density turbulence level, the mean and the maximum magnetic field values, the mean and the maximum density values, calculated over the whole simulation box at the final time in each simulation, are reported.}
\label{table:table1}
\end{table}

\begin{figure}[!htbp]
\centering
\includegraphics[width=1\textwidth]{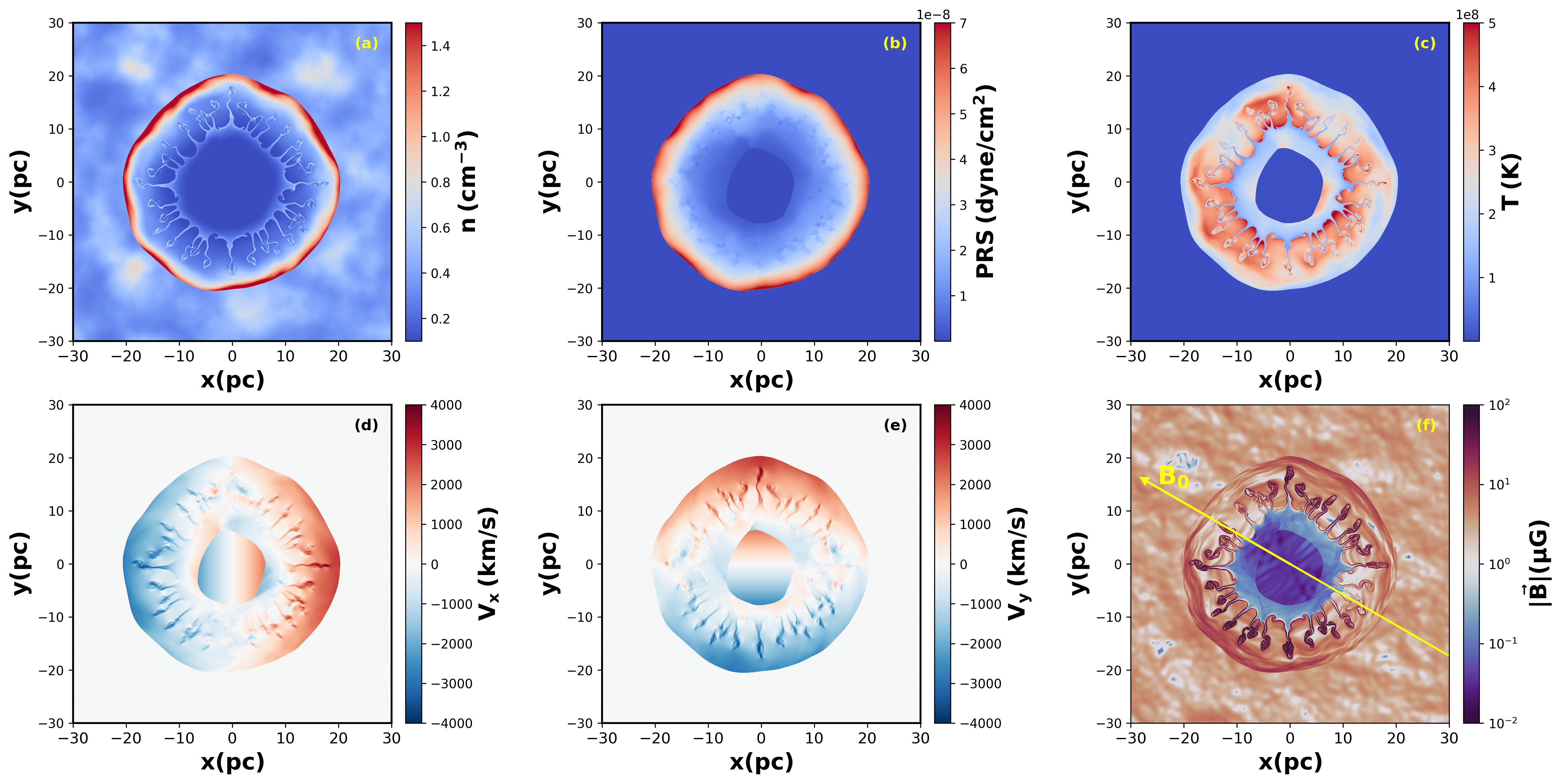}
\caption{Spatial distribution of number density (a), pressure (b), temperature (c), x- and y-components of the velocity (d,e), and the magnetic field magnitude (f) from Run 2 at an age of 3000 years.}
\label{fig:maps-sim}
\end{figure}

In Figure \ref{fig:maps-sim} we show for Run 2 at the final simulation time, $t=$ 3 $\times 10^3$ years, the spatial distributions of the number density, the kinetic pressure, the gas temperature, the velocity components, and the magnetic field intensity. At the beginning of the simulation, the ejecta is characterized by high values of density and pressure, such that a strong shock in the turbulent background is induced, heating and compressing the ambient medium. Since the shocked circumstellar medium pushes back the ejecta, a reverse shock (RS) is produced. In Figure \ref{fig:maps-sim} the RS extends up to about $10$ pc, while the forward shock (FS) expands beyond $20$ pc.

\begin{figure}[!htbp]
\centering
\includegraphics[width=1\textwidth]{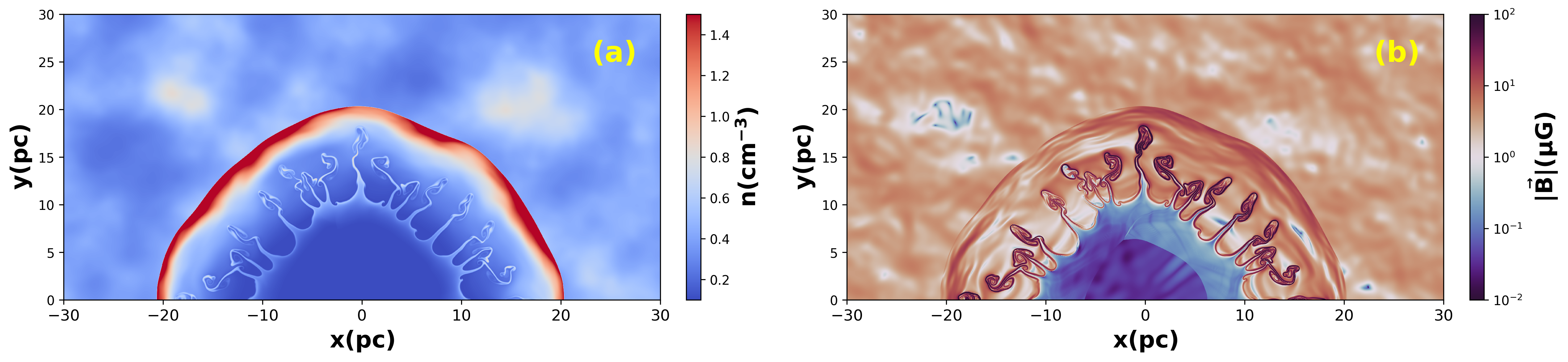}
\caption{Zoom of the spatial distributions of the number density (a) and of the magnetic field amplitude (b).}
\label{fig:maps-zoom}
\end{figure}

Between the FS and RS, the contact discontinuity (CD) emerges, where the RTI is triggered \citep{sharp1984overview}. Outside the inner ejecta, the region called the outer ejecta extends up to the FS, characterized by very high values of compression forms. The interaction between the expanding FS and the dense interstellar medium distorts the shock surface. To better characterize the FS, its magnetic compression ratio $r_B=B_d/B_u$, where $B_d$ is the downstream magnetic field and $B_u$ is the upstream magnetic field, has been calculated in the region in which the shock is quasi-perpendicular. We find that $r_B= 2.5$ is comparable with the ratio between the downstream ($n_d$) and the upstream ($n_{up}$) densities, namely $n_d/n_{up} = 2.4$, which is what is expected at perpendicular shocks.
Figures \ref{fig:maps-sim}(b) and (c) show the pressure and the temperature maps. The pressure map shows a behavior similar to the density, with higher values in the FS region. The temperature map shows the higher values in the region where the RT dominates.
Finally, the magnetic field exhibits a noticeable amplification and compression in the direction perpendicular to the direction of the mean magnetic field. The background magnetic field fluctuations tend to distort the shape of the FS and, as a result, it is possible to observe knots and filaments in this region, which reflect also in the non-homogeneous profile of kinetic pressure and temperature.
More details can be observed in Figure \ref{fig:maps-zoom}, where a zoom of the spatial distributions of the number density and magnetic field amplitude are presented. It is possible to see the structures formed by the triggered RTI and the deformation of the FS due to the interaction of the ejecta with the turbulent environment. The density map shows a more uniform compression in the FS region, while the magnetic field map shows a higher level of distortion in the shock surface and a magnetic field amplification in the regions where $\mathbf{B}$ is tangential, according to the Rankine-Hugoniot conditions. 
Notice that \citet{reynoso2013radio} present a radio polarization study of SN 1006 finding that particles are efficiently accelerated along the direction of the mean magnetic field (namely, in the northeast and southwest shells of the remnant). On the other hand, we perform purely MHD simulations. In our case, the magnetic field is amplified in the north-western part and the south-eastern part in agreement with the Rankine-Hugoniot jump conditions for an ideal MHD shock; indeed in such regions, the mean magnetic field is almost perpendicular to the shock normal, thus resulting in a higher B compression. In our simulations, we neglect the contribution of particle acceleration that plays a key role in magnetic field amplification \citep{bell2004turbulent}, a mechanism that will be explored in future works. Nevertheless, wherever emission is high, our analysis technique can shed light on the turbulence properties of the medium.

\begin{figure}[!htbp]
\centering
\includegraphics[width=1\textwidth]{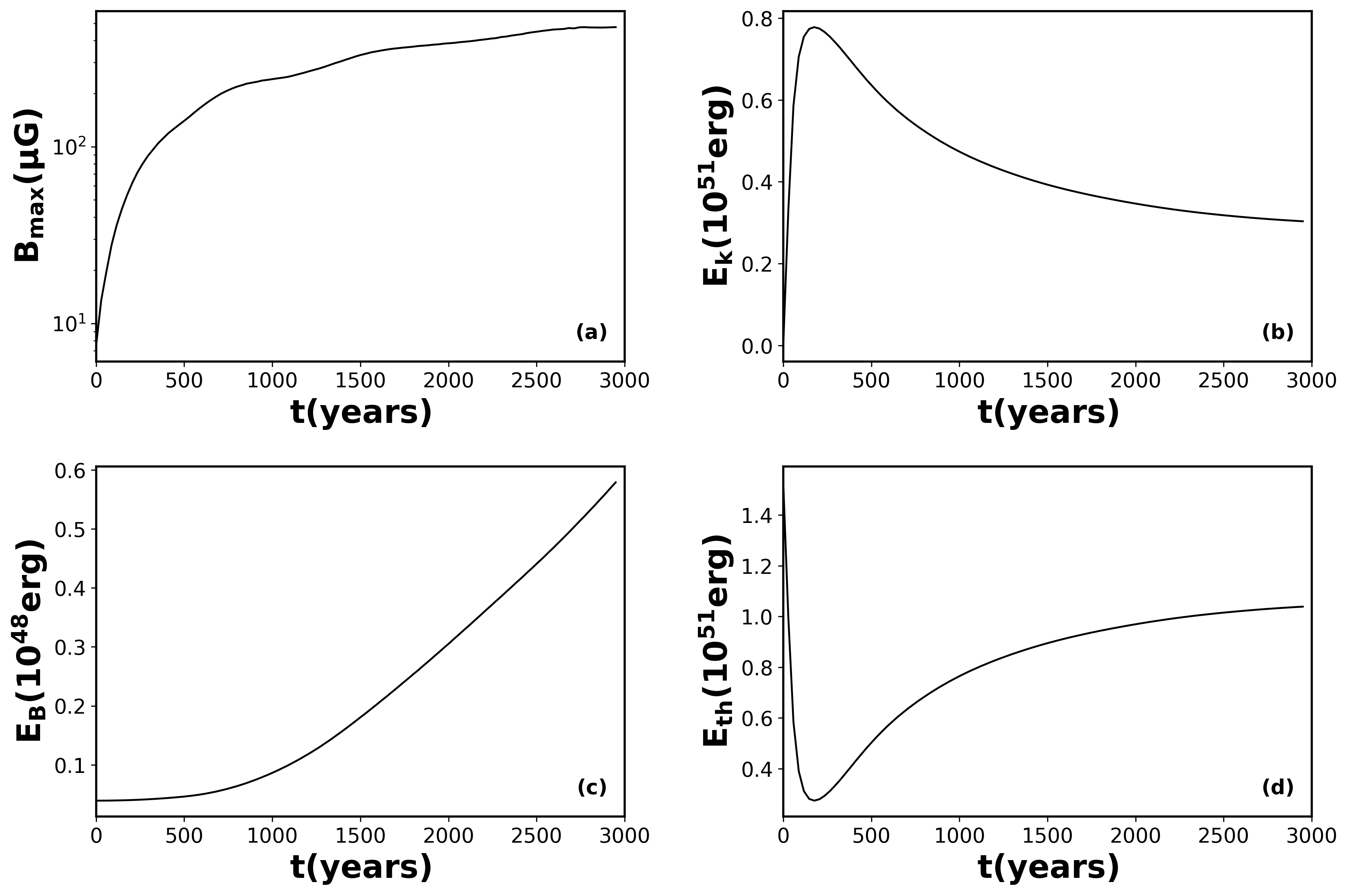}
\caption{(a) Time evolution of the maximum value of the magnetic field in the domain, (b) the total kinetic energy, (c) the magnetic energy and (d) the thermal energy, all averaged over the whole simulation domain.}
\label{fig:energies-dbB}
\end{figure}

In order to give an idea of the SNR evolution, we trace in Figure \ref{fig:energies-dbB} the time evolution of the maximum magnetic field attainable in the simulation, and the time behaviour of the kinetic, magnetic, and thermal energies. The energies are averaged over the whole simulation domain.

The maximum value of $\vert  \mathbf{B} \vert$  shows a linear increase until $\simeq$ 600 years in which it reaches 250 $\mu $G. After a brief period during which a little plateau appears (around 500 years), it rapidly increases until 3000 years and reaches a maximum value of 475 $\mu$G. 
The values obtained for the maximum amplification of the magnetic field strength are determined over the whole simulation box for each run (see Table \ref{table:table1}). We then calculate the mean magnetic field strength within the region between the RS and the FS. We find at 3$\times$ 10$^3$ years $B_{mean} \simeq$  13 $\mu$G and a maximum value of $B_{max} \simeq $ 475 $\mu$G, namely  the magnetic field amplification can reach up to almost two orders of magnitude with respect to the value of 3$\mu$G in the ISM. This result is in agreement with SN 1006 observations reported by \citep{Ressler2014}, in which the magnetic field value is of the order of 100 $\mu$G.

The behavior of the maximum value of the magnetic field is reflected on the magnetic energy that continuously increases, reaching a maximum value of $\sim$ $0.6 \times 10^{48}$ erg. These results are in agreement with the ones found by \citep{Giacalone_2007,guo2012amplification}.
Figures \ref{fig:energies-dbB} (b) and (d) show that the kinetic energy increases rapidly until 200 years due to the high density and pressure that drive the strong shock in the interstellar turbulent medium, while the thermal energy decreases rapidly with the same behaviour. The kinetic energy, during this rapid increase, reaches a maximum value of $\sim$ $0.8 \times 10^{51} $ erg that corresponds to a conversion of about 54\% of the thermal energy. After that, the ejecta slow down due to the interaction with the turbulent ISM so that the kinetic energy slowly decreases and the thermal energy increases. This happens because the total energy is mostly conserved and the magnetic energy remains much smaller than the other two.

\begin{figure}[!htbp]
\centering
\includegraphics[width=0.5\textwidth]{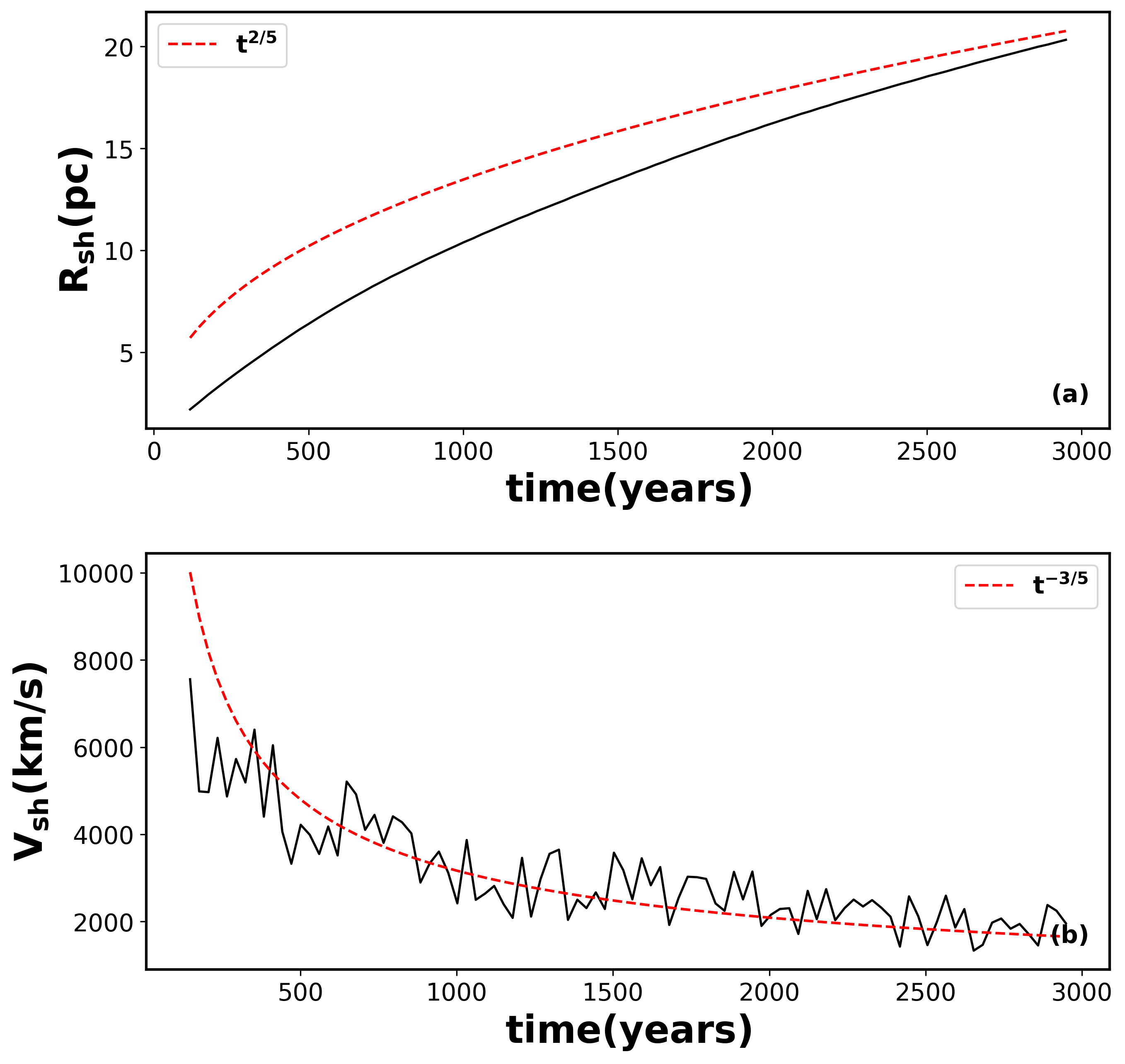}
\caption{Time evolution of the FS radius (a) and of the FS speed (b) in Run 2. A comparison with the trends predicted by the Sedov-Taylor phases are also reported (red dashed lines). }
\label{fig:R-Vsh-time}
\end{figure}

In Figure \ref{fig:R-Vsh-time} we show the time evolution of the SNR radius (a) and of the FS speed (b). Since the ejecta is expanding in a dense and turbulent environment, the shock velocity constantly decreases. The shock speed decreases from about 6000 km/s to 3000 km/s, in agreement with the theoretical evolution of the FS speed in the Sedov-Taylor phase, i.e. $V_{sh} \propto t^{-3/5}$ \citep{Helder_2012}. The radius of the FS also increases and approaches the power-law behaviour predicted in the Sedov-Taylor phase $r \propto t^{2/5}$ \citep{Chevalier82, Helder_2012, guo2012amplification}.

\subsection{Statistical analysis of turbulence at SNRs}
\begin{figure}[!htbp]
\centering
\includegraphics[width=1\textwidth]{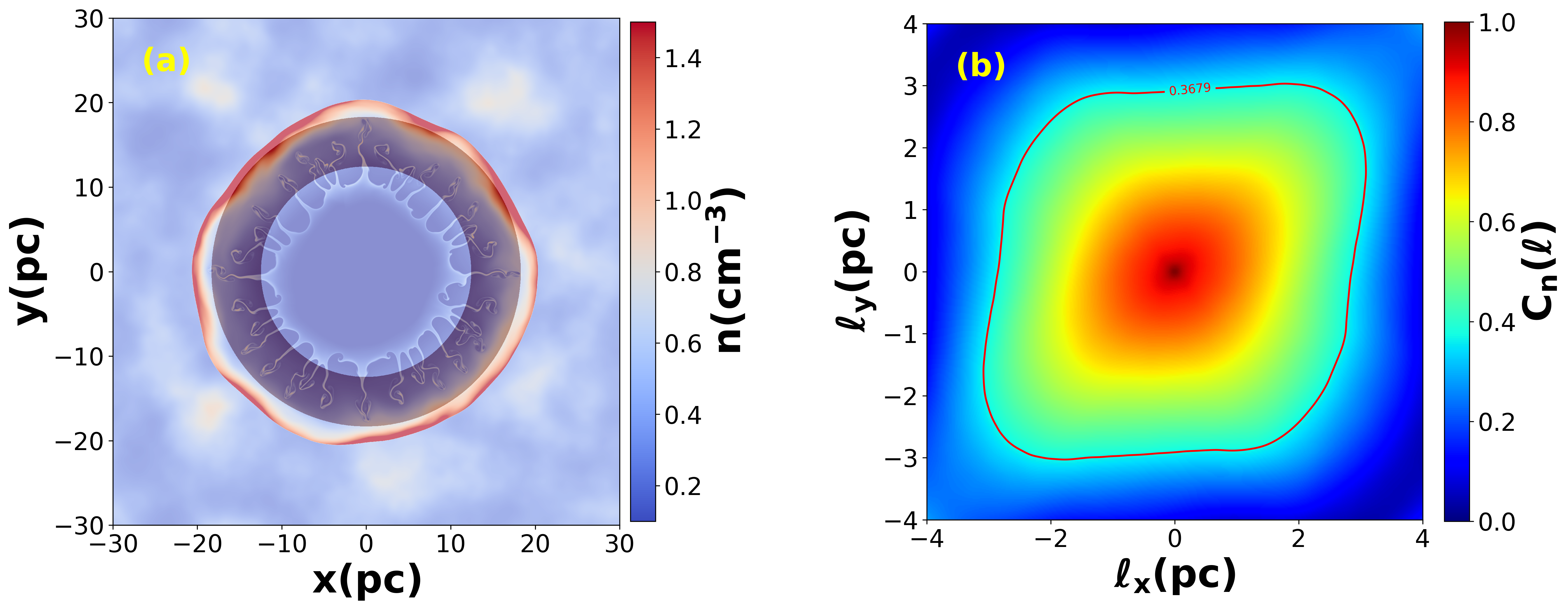}
\caption{(a) Density map distribution after the application of the ring-shaped mask (overplotted in gray), located within the CD. (b) Autocorrelation function map computed on the ring-shaped region. The red solid line represents the autocorrelation length, namely the isocontour of $C_2=1/e$. $\ell_x$ and $\ell_y$ are the increments along the $x$ and $y$ directions, respectively.}
\label{fig:maprho}
\end{figure}

\begin{figure}[!htbp]
\centering
\includegraphics[width=1\textwidth]{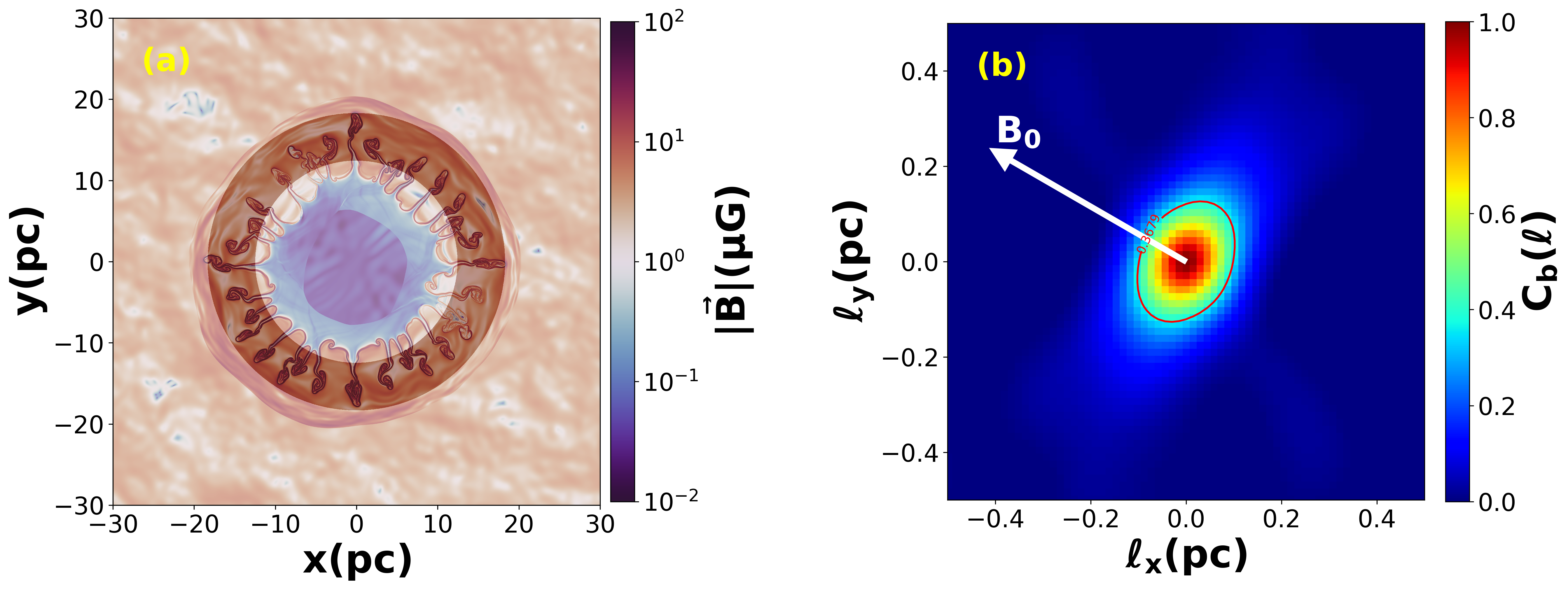}
\caption{Same as Figure \ref{fig:maprho} but for the magnetic field spatial map.}
\label{fig:mapB}
\end{figure}

We investigate the level of turbulence anisotropy related to the presence of a strong background magnetic field. \citet{shebalin1983anisotropy} found that, starting from an initial isotropic spectrum, the anisotropy develops also with a modest amplitude of the mean magnetic field. Different levels of anisotropies can also be found in different regions of the solar wind \citep{Dasso2005,weygand2009anisotropy,sahraoui2010three,chen2011anisotropy,bandyopadhyay2021geometry,pecora2023three}. To study turbulence properties, we use a novel technique: we apply a Heaviside mask (ring-shaped) on the density and the magnetic field maps, to select the region between the RS and the FS only. Thus, we calculate the 2D structure function $S_2(\Bell)$ of the magnetic field turbulence inside that ring mask, as

\begin{equation}
S_2(\Bell) =  \langle |\delta \bm{B}(\Bell+\mathbf{r}) - \delta \bm{B}(\mathbf{r})|^2 \rangle,
\label{eqn:autocorr}
\end{equation}

where $\delta\bm{B}(\bm{r}) = \bm{B}(\bm{r}) - \bm{B}_0$, $\bf{\Bell}$ represents a spatial lag, and $\bf{r}$ is the position in the simulation domain. Once we determine the structure function, we can calculate the autocorrelation function as $C_b(\Bell) = ( C_b(0) - S_2(\Bell)/2)/C_b(0)$. 
In order to get into deep statistical properties of the turbulence, we also calculate the correlation length $\lambda_c$. The correlation length is defined as the e-folding distance of the normalized autocorrelation function \citep{Smith2001}. It gives us a measure of the size of the turbulence structures.
The same approach has been adopted for the density field. In Figures \ref{fig:maprho} and \ref{fig:mapB} the outputs of the above analysis are displayed, for the density map and the magnetic field map, respectively.

From Figure \ref{fig:maprho} the autocorrelation density map appears to be anisotropic, elongated along the direction perpendicular to the mean magnetic field. A similar result has been observed for the magnetic field map, where perpendicularly to the mean magnetic field the largest values of the correlation length are detected. 

In addition, the density field tends to be more correlated over larger spatial scales.
In both cases, the explanation comes from the structures emerging in the CD due to the RT instability.

\subsubsection{Simulations with different background turbulence levels} \label{sec:diff-bckg}
We now discuss how the variation of the parameters of the simulations reported in Table \ref{table:table1} affects the evolution of the SNR. 

\begin{figure}[!htbp]
\centering
\includegraphics[width=1\textwidth]{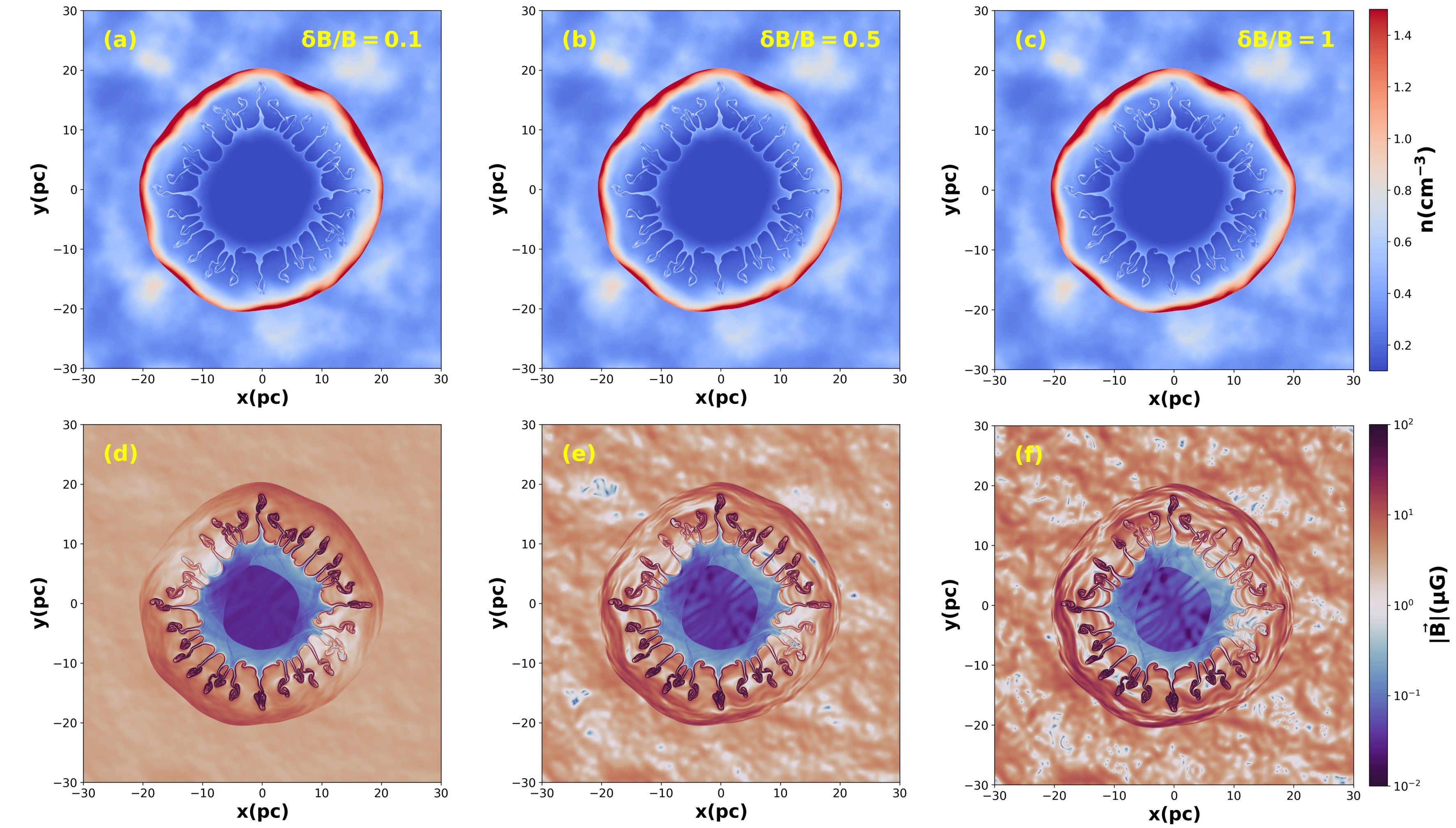}
\caption{Spatial distribution of the number density((a),(b),(c)) and of the magnetic field magnitude ((d),(e),(f)) from the 2D simulation at an age of 3000 years in Run 1, Run 2 and Run 3. }
\label{fig:dbB-var}
\end{figure}

\begin{figure}[!htbp]
\centering
\includegraphics[width=1\textwidth]{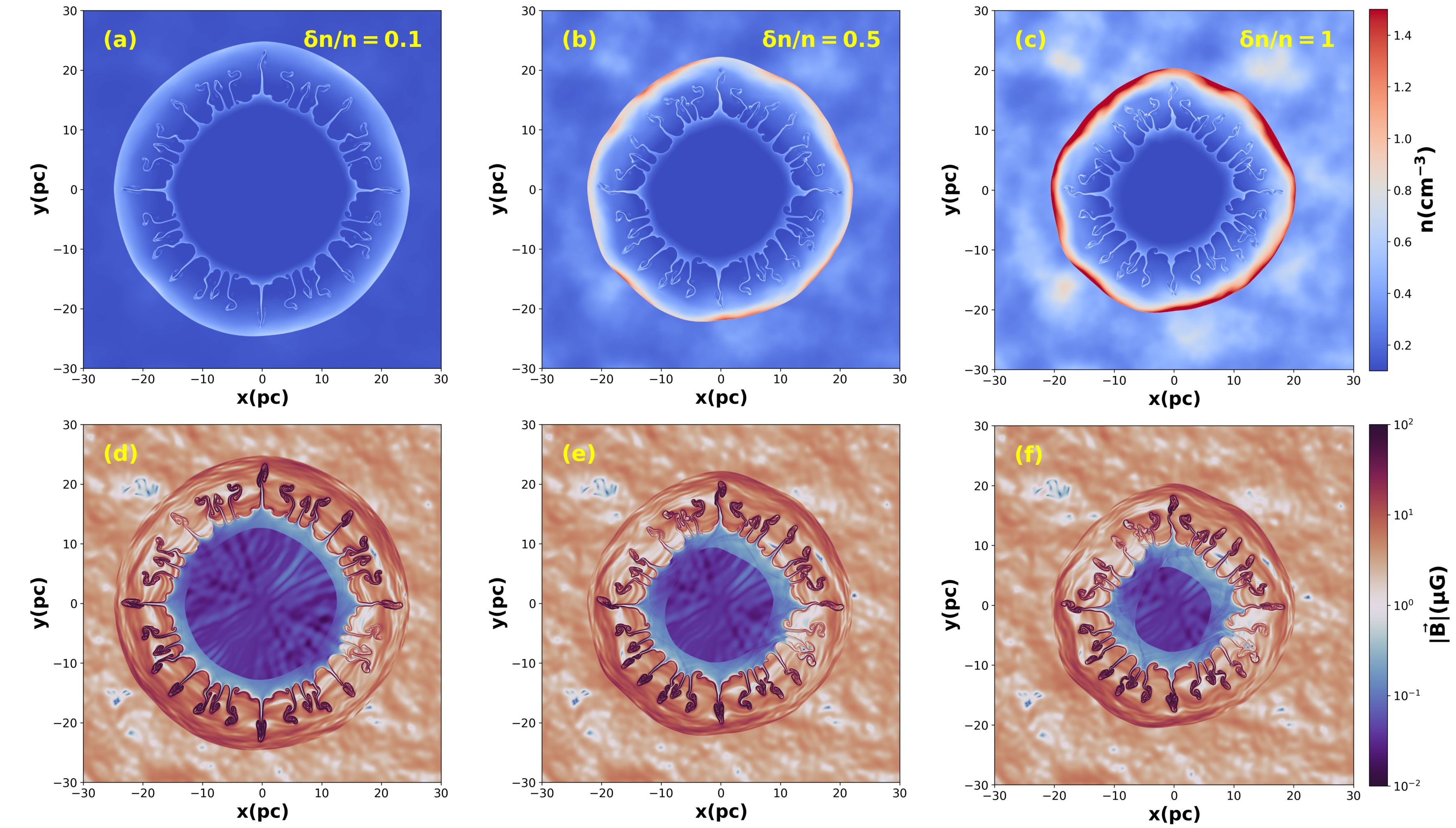}
\caption{Spatial distribution of the number density \textbf{[(a)-(c)]} and of  the magnetic field magnitude \textbf{[(d)-(f)]} from the 2D simulation at an age of 3000 years in Run4, Run5 and Run2. }
\label{fig:dnN-var}
\end{figure}

Figure \ref{fig:dbB-var} compares the spatial maps of the density and of the magnetic field in Run 1-3, where the level of magnetic field turbulence varies and the density fluctuations are fixed. The density field does not exhibit a significant variation with the $\delta \bm{B}/\bm{B}$ parameter, while the magnetic field in the SNR reacts to the amplitude of the ISM perturbed field in terms of the formation of knots and filaments. Furthermore, as expected, the increase of the magnetic field turbulence leads to an enhancement of the magnetic field magnitude along the shock. \\
In Figure \ref{fig:dnN-var} results from Run 2, 4, and 5 are presented. Unlike the previous case, as the $\delta n/n$ increases, the density maps show a stronger deformation of the shock surface. As a consequence, the SNR radius decreases. In the case of the magnetic field maps (Figure \ref{fig:dnN-var} (d),(e),(f)) we do not observe variation in terms of magnetic field amplification, but the only difference is again related to the radius of the remnant. We can conclude that when we set $\delta n/n$ to a small value, even if we vary $\delta \bf{B}/\bf{B}$, the rest expand in the same way in the three cases. Instead, when we fix $\delta \bf{B}/\bf{B}$ and vary $\delta n/n$, the expansion is slowed down.



\begin{figure}[!htbp]
\centering
\includegraphics[width=1\textwidth]{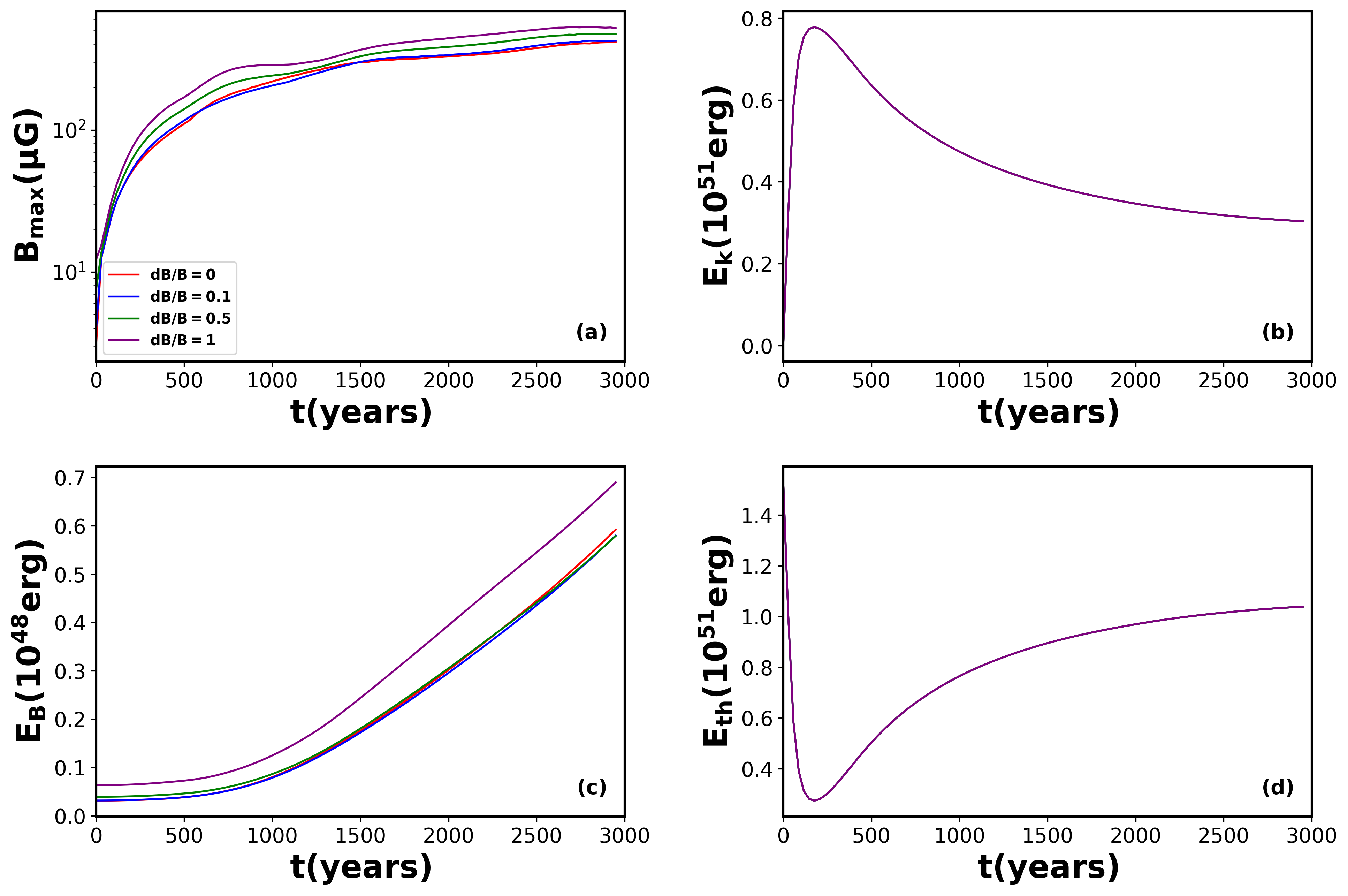}
\caption{Same quantities as in Figure \ref{fig:energies-dbB}. The blue, green, and purple lines indicate Run 1, Run 2 and Run 3, respectively. The red line represents the non-turbulent case.}
\label{fig:energies-dbB-comparison}
\end{figure}

In Figure \ref{fig:energies-dbB-comparison} the averaged energies and the maximum magnetic field time behavior over the three simulations are displayed. For comparison, the red line in each panel represents the simulation with a uniform and non turbulent ISM. The magnetic energy and the maximum value of $\vert \bm{B}\vert$ reach higher values as the amplitude of the magnetic field fluctuations increases (panels (a) and (c) in Figure \ref{fig:energies-dbB-comparison}).

The values observed in Figure \ref{fig:energies-dbB-comparison} are in agreement with the results obtained by \citet{guo2012amplification} and \citet{xu2017magnetic}. In particular, the values of $B_{max}$ in our simulations at  t=$3 \times 10^3$ years are approaching the mG value, similar to the values reported in \citet{xu2017magnetic}.

The kinetic and the thermal energies do not show variations. This is due to the lower increase in the level of density fluctuations. A similar study has been performed by fixing $\delta \bm{B}/\bm{B}$ and varying $\delta n/n$ (not shown).

\begin{figure}[!htbp]
\centering
\includegraphics[width=1\textwidth]{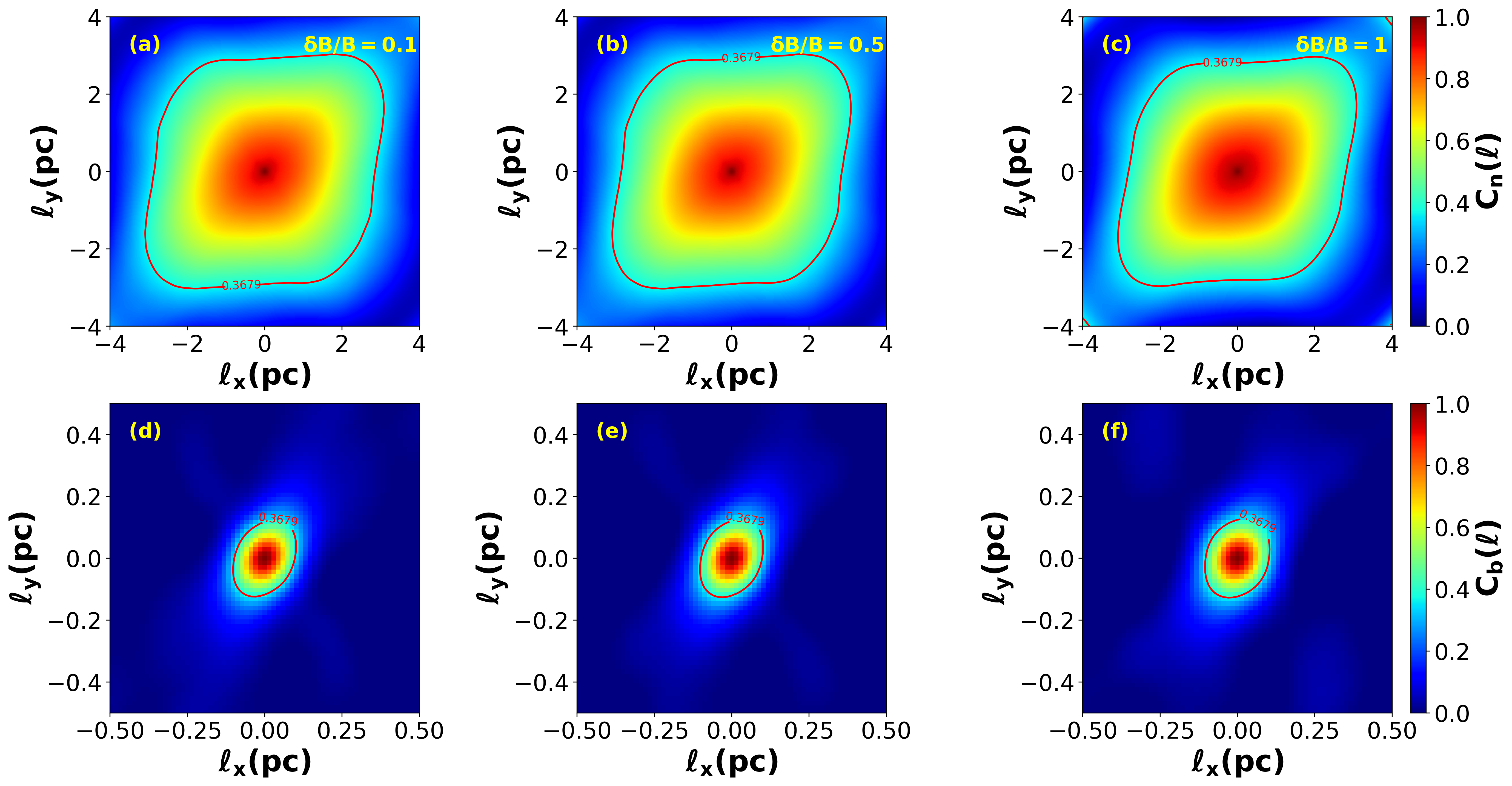}
\caption{Same as Figure \ref{fig:mapB}, but for Run1, Run2 and Run3.}
\label{fig:autocorr-dbB}
\end{figure}

\begin{figure}[!htbp]
\centering
\includegraphics[width=1\textwidth]{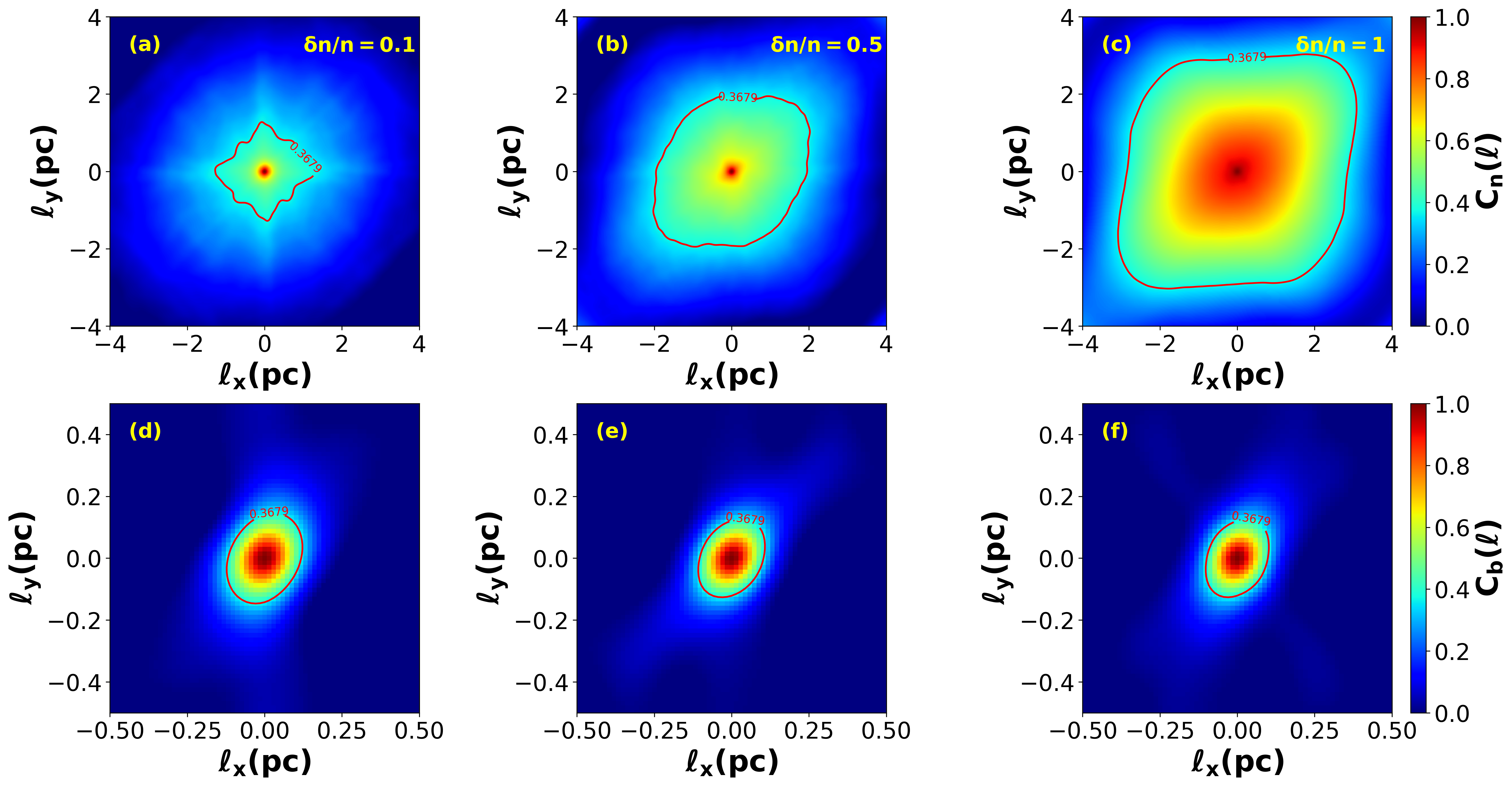}
\caption{Same as Figure \ref{fig:mapB}, but for Run 4, Run 5 and Run 2. }
\label{fig:autocorr-dnN}
\end{figure}

The behavior of the magnetic field strength 
is similar in the three cases studied, with lower values of $B_{max}$ for $\delta n/n=1$. This is due to the fact that the expansion of the remnant is slowed down by the large amplitude of the density fluctuations. Indeed, the magnetic energy undergoes a slower increase as $\delta n/n$ increases. 
As a consequence, the kinetic energy tends to be higher for $\delta n/n =0.1$ and the thermal energy presents an opposite behavior. 
Finally, as we made in Section \ref{sec:numerical results}, using the ring-shaped mask, we determine the autocorrelation function maps for all the cases analyzed, to make a comparison between the density and magnetic field anisotropy level. We report the comparison among all the simulations in Figure \ref{fig:autocorr-dbB} and \ref{fig:autocorr-dnN}. Figures \ref{fig:autocorr-dbB} (a), (b), and (c) show the autocorrelation function applied on the density field for Run 1, Run 2, and Run 3. Figures show that there are no ``significant'' differences among the three maps, with results similar to those observed in Figure \ref{fig:maprho}. In the case of Figures \ref{fig:autocorr-dbB} (d), (e) and (f) the maps are similar between them, although it is expected a higher degree of anisotropy in the case of low $\delta \bm{B}/\bm{B}$, and a more isotropic distribution for high $\delta \bm{B}/\bm{B}$. This is due to the choice of the mask, as seen in Figure \ref{fig:mapB}, which is taken inside the CD, where the RT instability dominates. The behavior expected in the autocorrelation maps is also observed in Figure \ref{fig:dbB-var} (d), (e), and (f), where the highest values of the magnetic field strength are distributed more uniformly in the region between the FS and the RS of the SNR when $\delta B/B$ reaches values equal (or similar) to 1. On the other hand, in Figure \ref{fig:autocorr-dnN} for $\delta n/n=0.1$ the autocorrelation map looks isotropic while when $\delta n/n$ starts to increase, the degree of anisotropy increases as well within the CD.


In order to determine quantitatively the degree of anisotropy observed in the autocorrelation maps, we calculate for each run the Shebalin's angle $\theta_{Sheb}$ \citep{shebalin1983anisotropy, Milano2001, Coscarella2020}. This angle was introduced by \citet{shebalin1983anisotropy} to measure the level of anisotropy in the k-spectrum for the case of magnetized plasma turbulence. The Shebalin's angle can be determined using the structure functions, obtained with the same technique used for the development of the autocorrelation map, thus

\[
\theta_{Sheb} = \arctan \biggl( \sqrt{ 2\lim_{\ell \to 0}  \frac{S_\perp(\ell)}{S_{//}(\ell)}} \biggr),
\]

where $S_\perp(\ell)$ and $S_{//}(\ell)$ indicate the structure functions in the perpendicular and parallel directions to the local mean magnetic field, respectively. In 3D simulations the isotropy is recovered when the values of $\theta_{Sheb}$ = $\arctan (\sqrt{2})$ = 54.74$^{\circ}$, while for higher values of the angle the level of anisotropy increases \citep{Milano2001}. In 2D simulations, the above expression is without the value 2, i.e. $tg^2(\theta_{Sheb}) = \lim_{\ell \to 0} \frac{S_\perp(\ell)}{S_{//}(\ell)}$. So the isotropy is observed when $\theta_{Sheb} = \arctan(1) = 45$$^{\circ}$. We calculate this angle in our simulations both for the magnetic and for the density fields, within the ring masks. The results are shown in Table \ref{table:table2}.

\begin{table}[!htbp]
\centering
\begin{tabular}{c c c c c }
\hline
Run &$  \text{\boldmath$\delta$} \bf{B/B}$ & $ \text{$\delta$} n/n$ & $\theta_{Sheb}(B)$ & $\theta_{Sheb}(n)$ \\ 
\hline
1 & 0.1 & 1 & 39.76 & 43.64 \\
2 & 0.5& 1 & 40.07 & 43.79 \\
3 & 1 &  1 & 41.25 & 43.84  \\
4 & 0.5 &  0.1 & 40.04 & 43.82  \\ 
5 & 0.5 &  0.5 & 39.79 & 43.25  \\ 

\hline
\end{tabular}
\caption{ Shebalin's angle determined for Run 1, 2, 3, 4 and 5.  }
\label{table:table2}
\end{table}

From the values presented in Table \ref{table:table2}, it appears evident, in both cases, that there is a certain degree of anisotropy since the values range between 39$^{\circ}$ and 43$^{\circ}$. In Run 1, 2 and 3, it seems that the level of anisotropy decreases as the level of magnetic turbulence of the ISM grows, while in the case of the density, the value of the angle is almost constant. 
Notice that the level of anisotropy is computed within a mask that covers the region in which the RTI dominates (see Figures \ref{fig:maprho} and \ref{fig:mapB}).
Increasing the level of magnetic fluctuations, the compression at shocks rises and, consequently, the RTI becomes more intense as seen in Figure \ref{fig:dbB-var} and the turbulence regime moves towards a more isotropic behavior. This could explain why we observe values of the Shebalin angle greater for higher $\delta B/B$. The orientation in the autocorrelation function map is related to the fact that the RTI keeps track of the mean magnetic field orientation and since the regions with higher compression are the regions perpendicular to the direction of the mean magnetic field the autocorrelation function decays slower along the perpendicular direction. Run 4 and 5 seem to show different behavior in terms of angle values, this happens since for very low values of $\delta n/n$, the distribution of density within the CD is more uniform than when $\delta n/n$ increases (see Figure \ref{fig:dnN-var}). 

\begin{figure}[!htbp]
\centering
\includegraphics[width=1\textwidth]{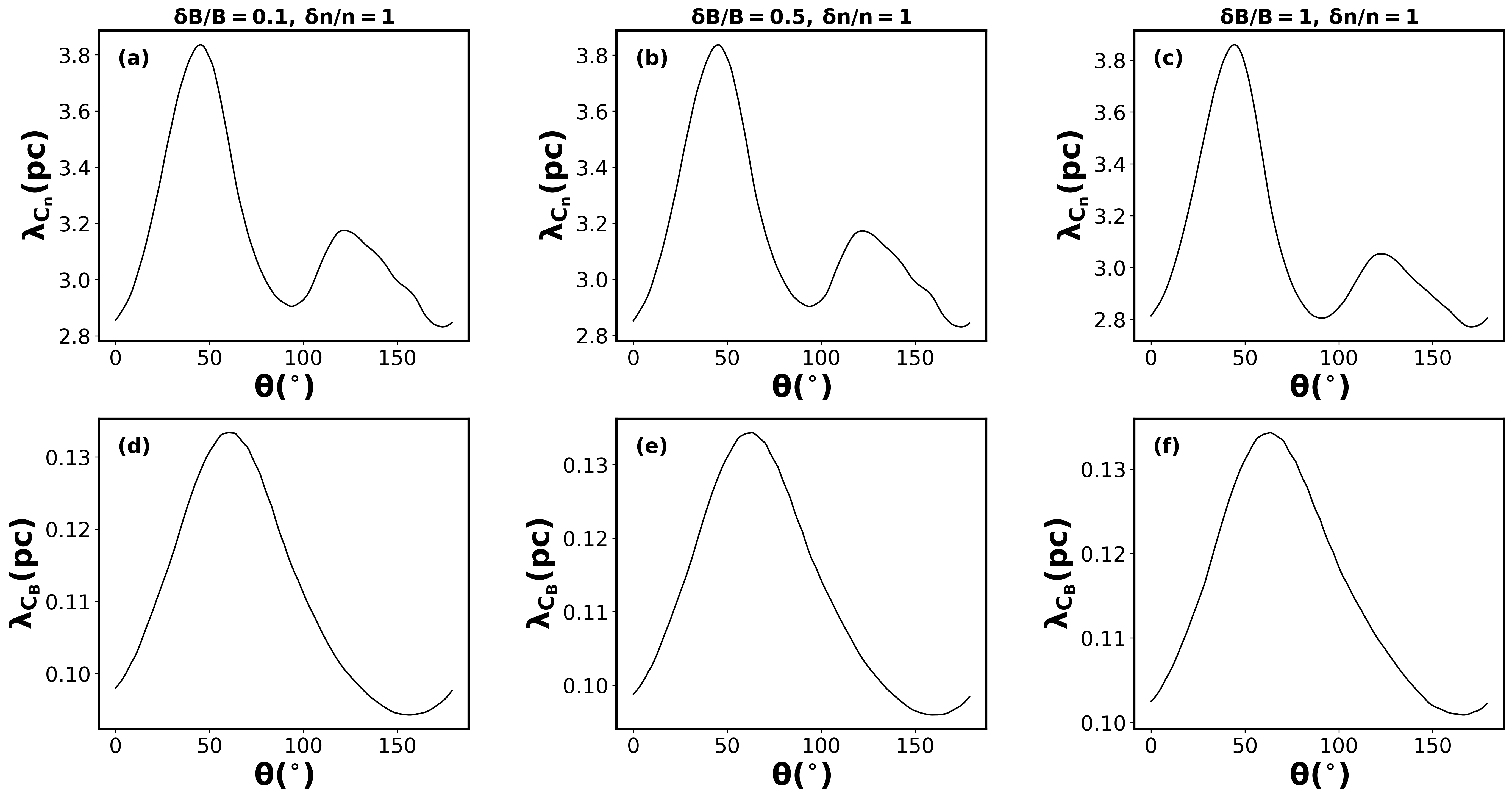}
\caption{Autocorrelation length as a function of angle obtained for Run 1,2,3.}
\label{fig:corrllen-dbB}
\end{figure}

Figure \ref{fig:corrllen-dbB} shows the correlation length obtained interpolating the autocorrelation function in polar coordinates $C_b(r,\theta)$. The different panels show the correlation length as a function of the angle, which ranges between 0$^{\circ}$ and 180$^{\circ}$.  Indeed,  we are considering the half-plane of the positive $y$-axis. Figure \ref{fig:corrllen-dbB} (a), (b), (c) show similar behavior, with two bumps that are around 50$^{\circ}$ and 130$^{\circ}$ (namely the directions almost perpendicular and parallel to the mean magnetic field), and a minimum value around 90$^{\circ}$ (the direction oblique to the mean magnetic field).

The difference between the two values is $\Delta \lambda = \lambda(50^{\circ}) - \lambda(130^{\circ})$ = 1.0 pc. In Figure \ref{fig:corrllen-dbB} (d), (e), (f), the correlation length of the magnetic field is reported. In this case we observe a maximum of the correlation length of the magnetic field around 60$^{\circ}$ (the direction perpendicular to the mean magnetic field) and a minimum at 150$^{\circ}$ (the direction parallel to the mean magnetic field). We calculate the difference between the maximum and the minimum values $\Delta \lambda $= 0.035 pc.

\begin{figure}[!htbp]
\centering
\includegraphics[width=1\textwidth]{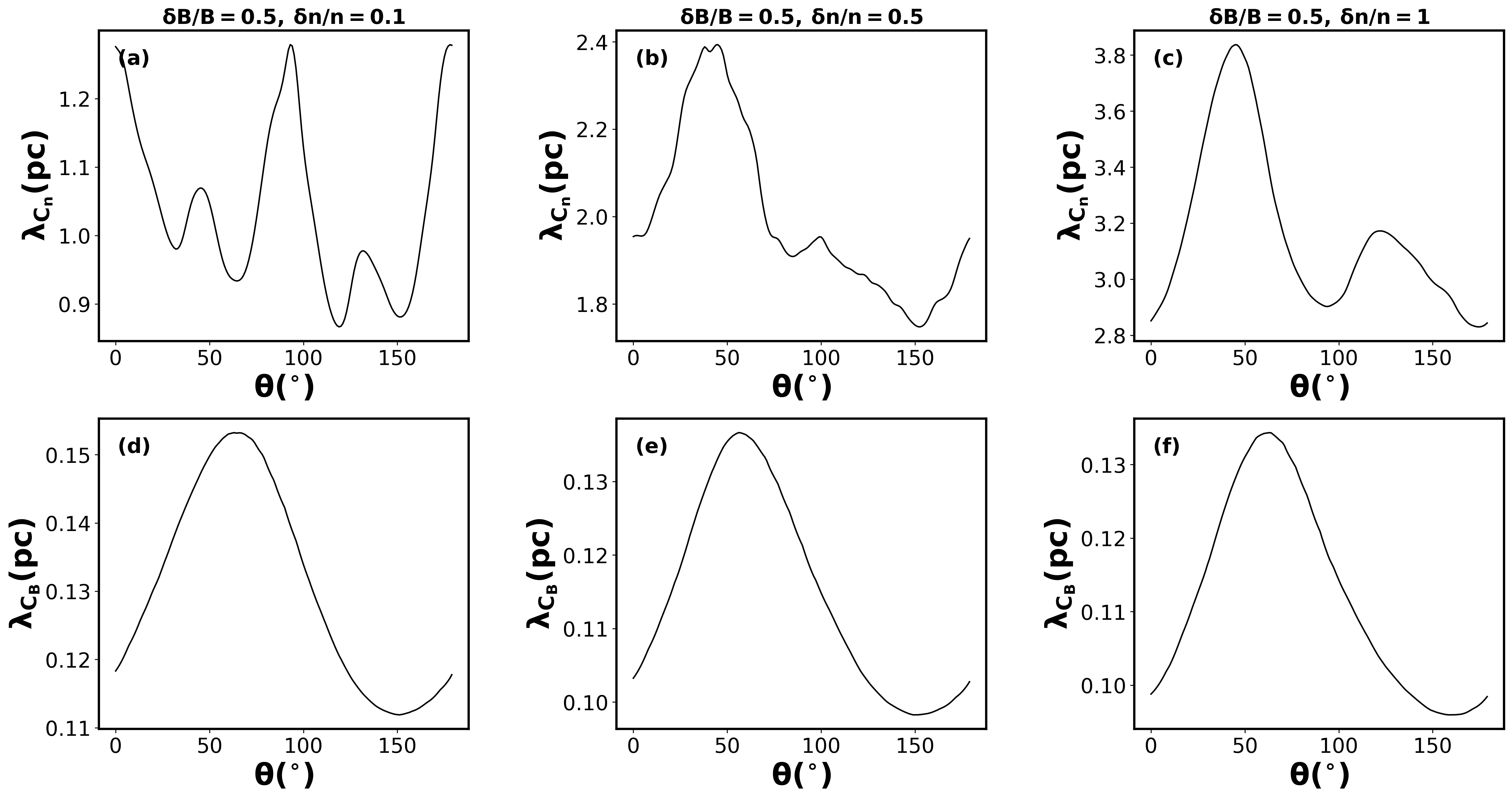}
\caption{Autocorrelation length as a function of angle obtained for Run 4,5,2.}
\label{fig:corrllen-dnN}
\end{figure}

Figure \ref{fig:corrllen-dnN} shows a behavior similar to Figure \ref{fig:corrllen-dbB} in the case of the autocorrelation length associated with the magnetic field, but differences appear for the same quantity calculated on the density. As we discussed above, the Rayleigh Taylor is more effective when the values of $\delta n/n$ are higher, so in the case of Figure \ref{fig:corrllen-dnN} (a) and (b) we can see a weaker degree of anisotropy ($\Delta \lambda_{C_{n}}=0.2$ pc in the case of panel (a) and $\Delta \lambda_{C_{n}}=0.6$ pc in the case of panel (b)) than in panel (c), with a variation of correlation length that is 1.0 pc.  

\begin{figure}[!htbp]
\centering
\includegraphics[width=0.5\textwidth]{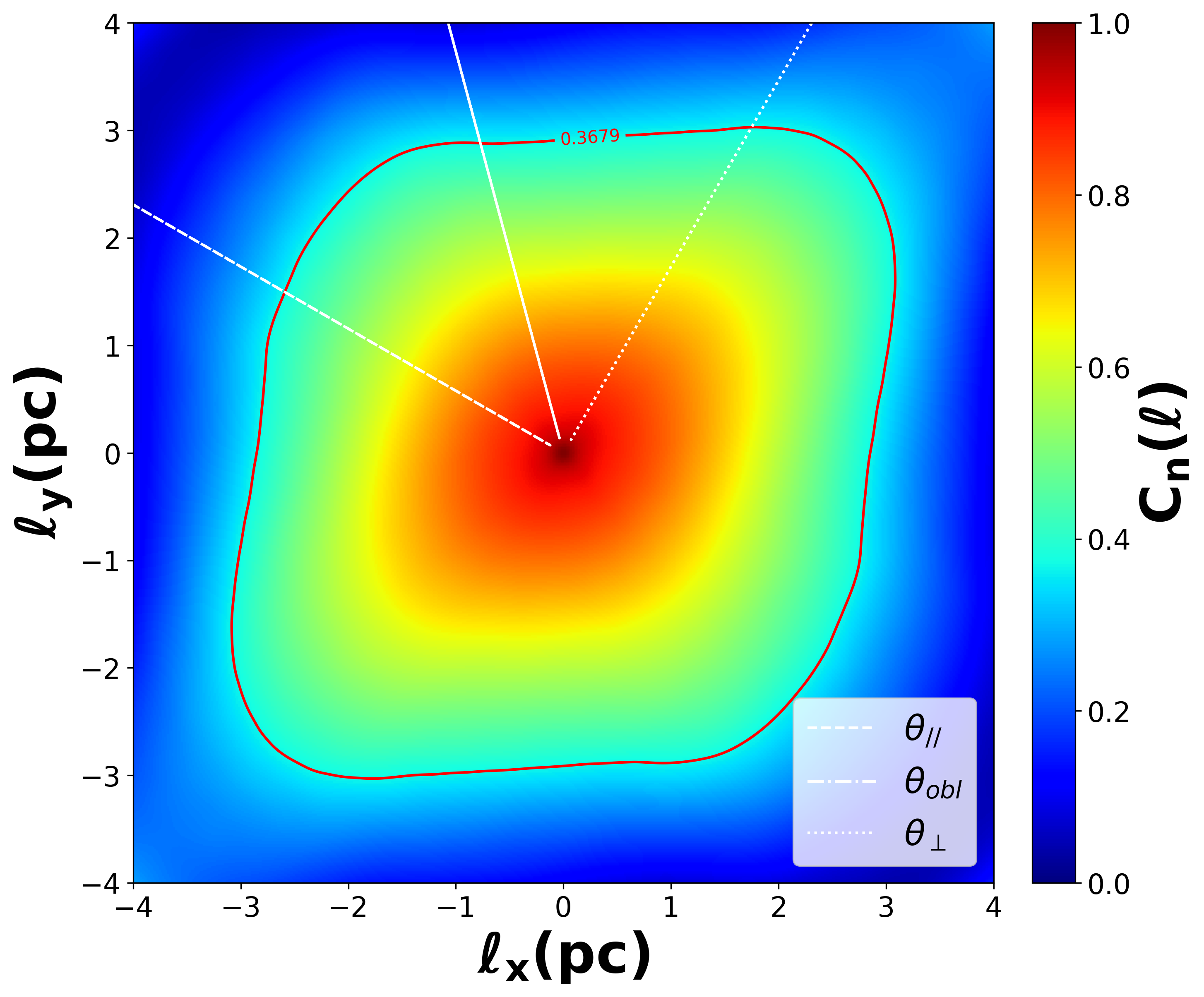}
\caption{Same Figure as Figure \ref{fig:maprho}(b). The solid white line indicates the mean magnetic field direction, while the dotted, dashed, and dot-dashed white lines represent the cuts made at three different angles (also indicated in the plot) with respect to $\mathbf{B_0}$. }
\label{fig:cuts-angle}
\end{figure}

Further, we have extracted a spatial spectrum making three cuts in the autocorrelation maps at different $\theta$ angles with respect to the mean magnetic field direction, as displayed in Figure \ref{fig:cuts-angle}. We have then computed the power spectral density along such directions using the Blackman-Tuckey technique. We define the spectra as
\[
P(k,\theta) = \int C_b(r,\theta) W(r) e^{-ikr} dr
\]
where $W(r)$ is the Hann function. We compare all these spectra with an omnidirectional spectrum. We choose directions \textit{a)} aligned with $\mathbf{B}_0$, \textit{b)} quasi perpendicular to it, and \textit{c)} perpendicular to $\mathbf{B}_0$.

\begin{figure}[!htbp]
\centering
\includegraphics[width=1\textwidth]{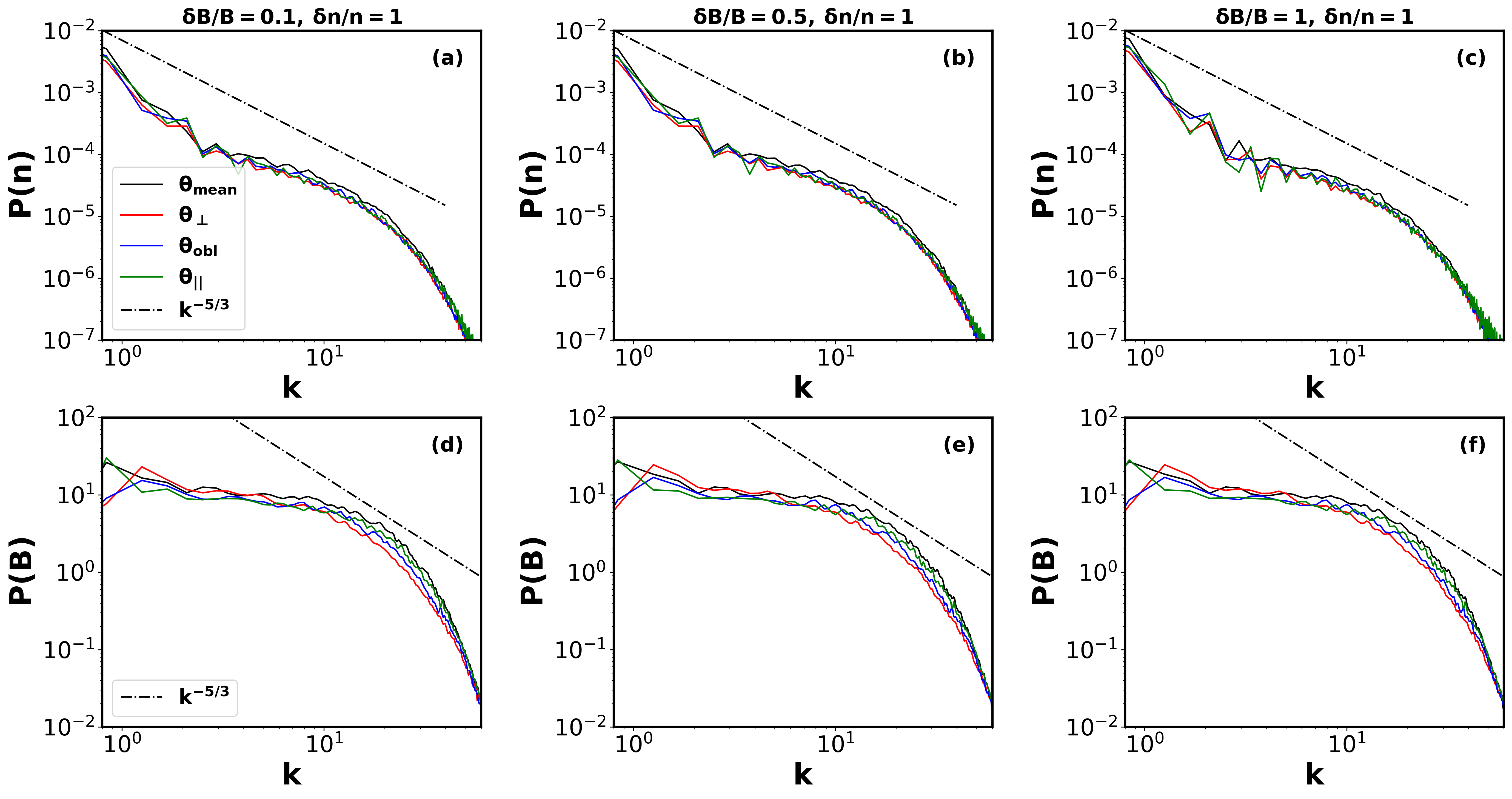}
\caption{Spectra obtained for Run1, Run2 and Run 3.}
\label{fig:spec-vardbB}
\end{figure}

\begin{figure}[!htbp]
\centering
\includegraphics[width=1\textwidth]{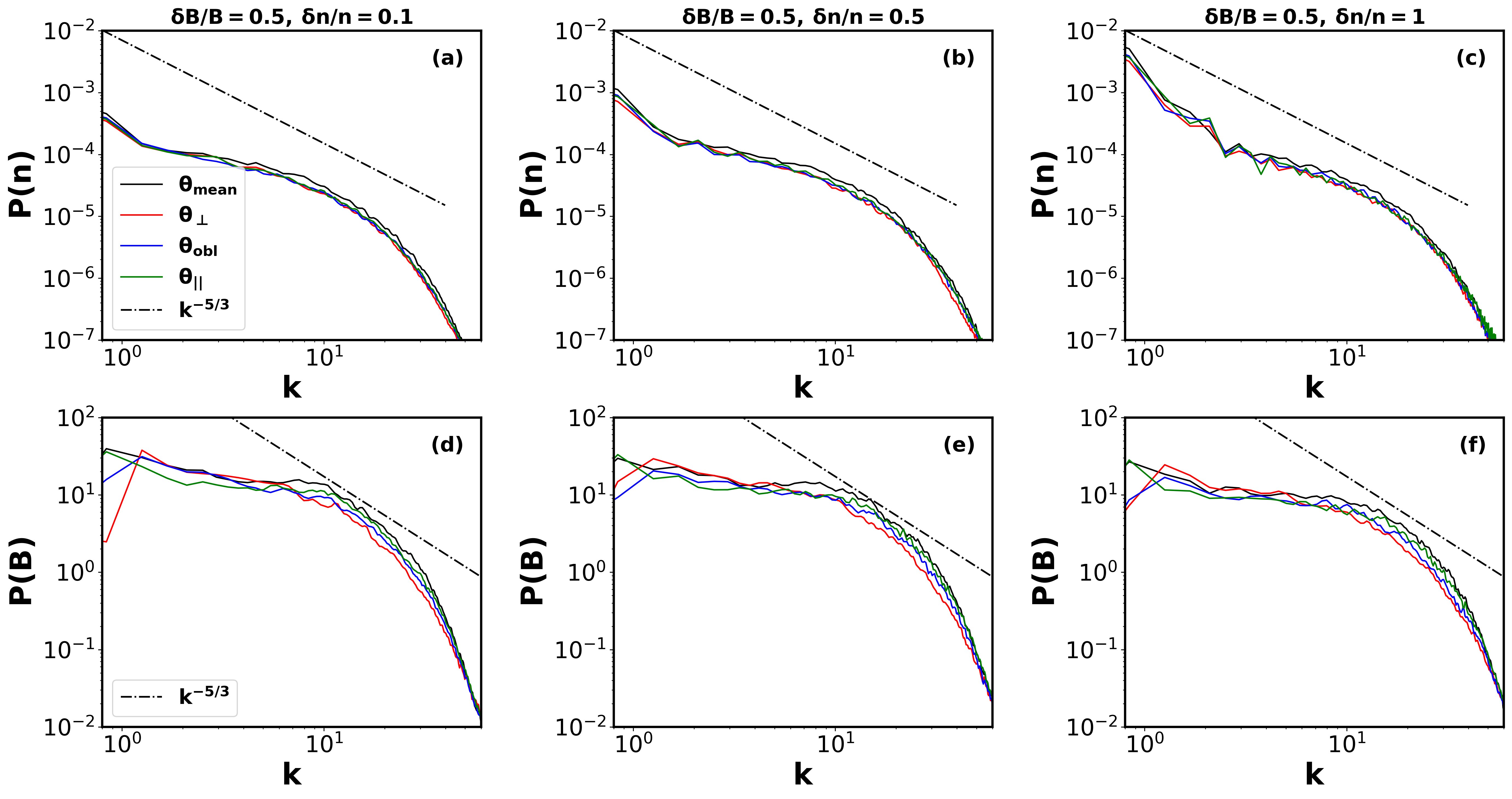}
\caption{Spectra obtained for Run4, Run5 and Run 2. }
\label{fig:spec-vardnN}
\end{figure}

The resulting power spectral densities for Run 1, Run 2, and Run 3 are reported in Figure \ref{fig:spec-vardbB} and for Run 4, Run 5, and Run 2 in Figure \ref{fig:spec-vardnN}. 

Figures \ref{fig:spec-vardbB} and \ref{fig:spec-vardnN} show analogous results. The density power spectra exhibit a Kolmogorov-type spectrum and it seems that there is no particular anisotropy because the spectra are completely overlapped. 
Magnetic field spectra show a different behavior. In order to understand if we are in a regime of fully developed turbulence we determined the eddy turnover time  $t_{eddy}=\lambda_{C_{b}}/V_{rms} \simeq $ 60 years. This implies that the regime of magnetic fully developed turbulence has been reached.
This may be related again to the presence of the RTI. Furthermore, the spectra of the three cuts show a slight anisotropic behavior as can be seen from the bottom panels in Figures \ref{fig:spec-vardbB} and \ref{fig:spec-vardnN}.

\subsection{Simulations with different $M_{ej}$}
\begin{figure}[!htbp]
\centering
\includegraphics[width=1\textwidth]{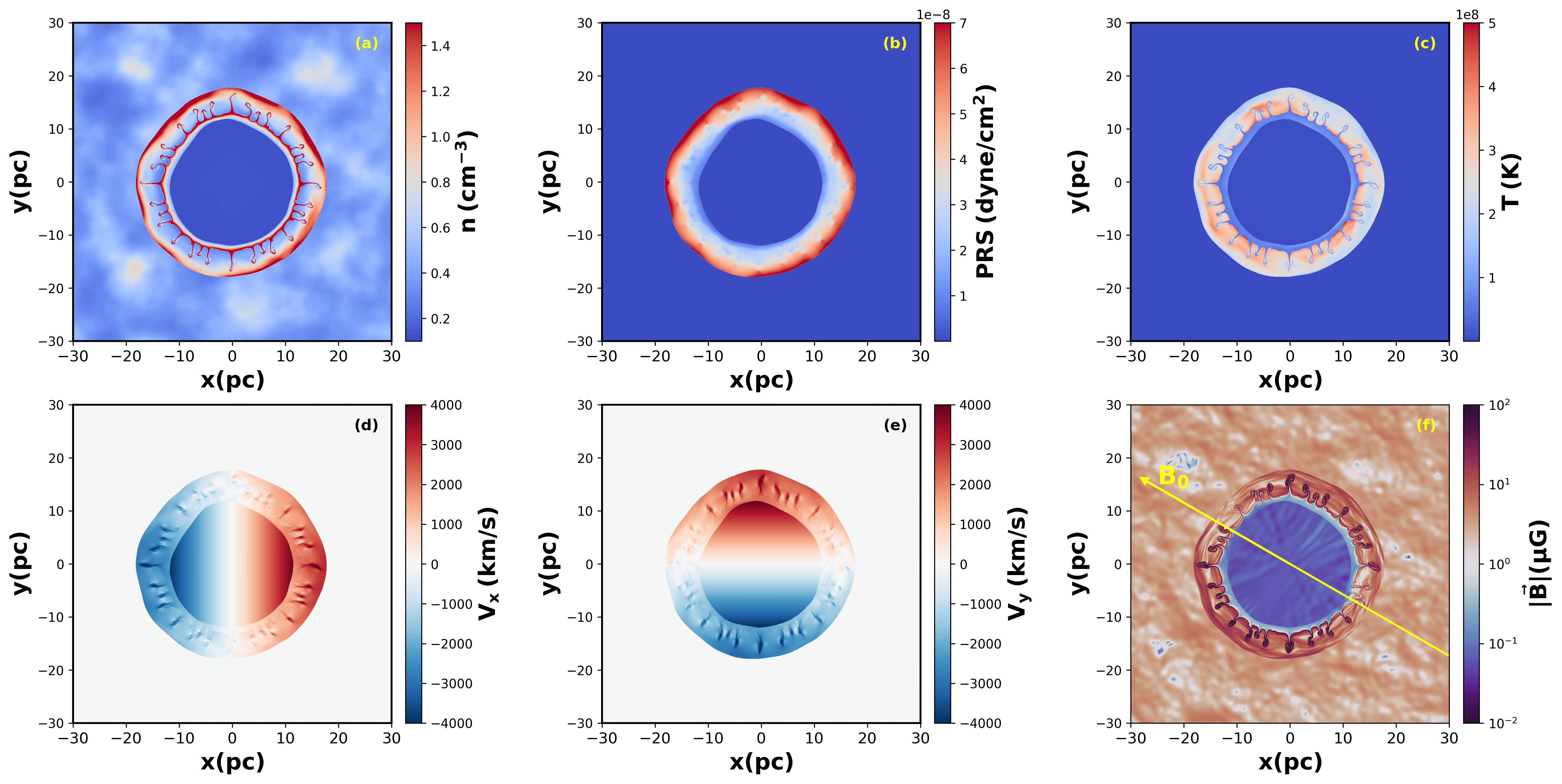}
\caption{Spatial distribution of the number density, pressure, temperature, x- and y-components of the velocity, and of the magnetic field magnitude from the 2D simulation with an age of 3000 years for the case $M_{ej}=3 M_{\odot}$ (Run 6). }
\label{fig:maps-3Mej}
\end{figure}

\begin{figure}[!htbp]
\centering
\includegraphics[width=1\textwidth]{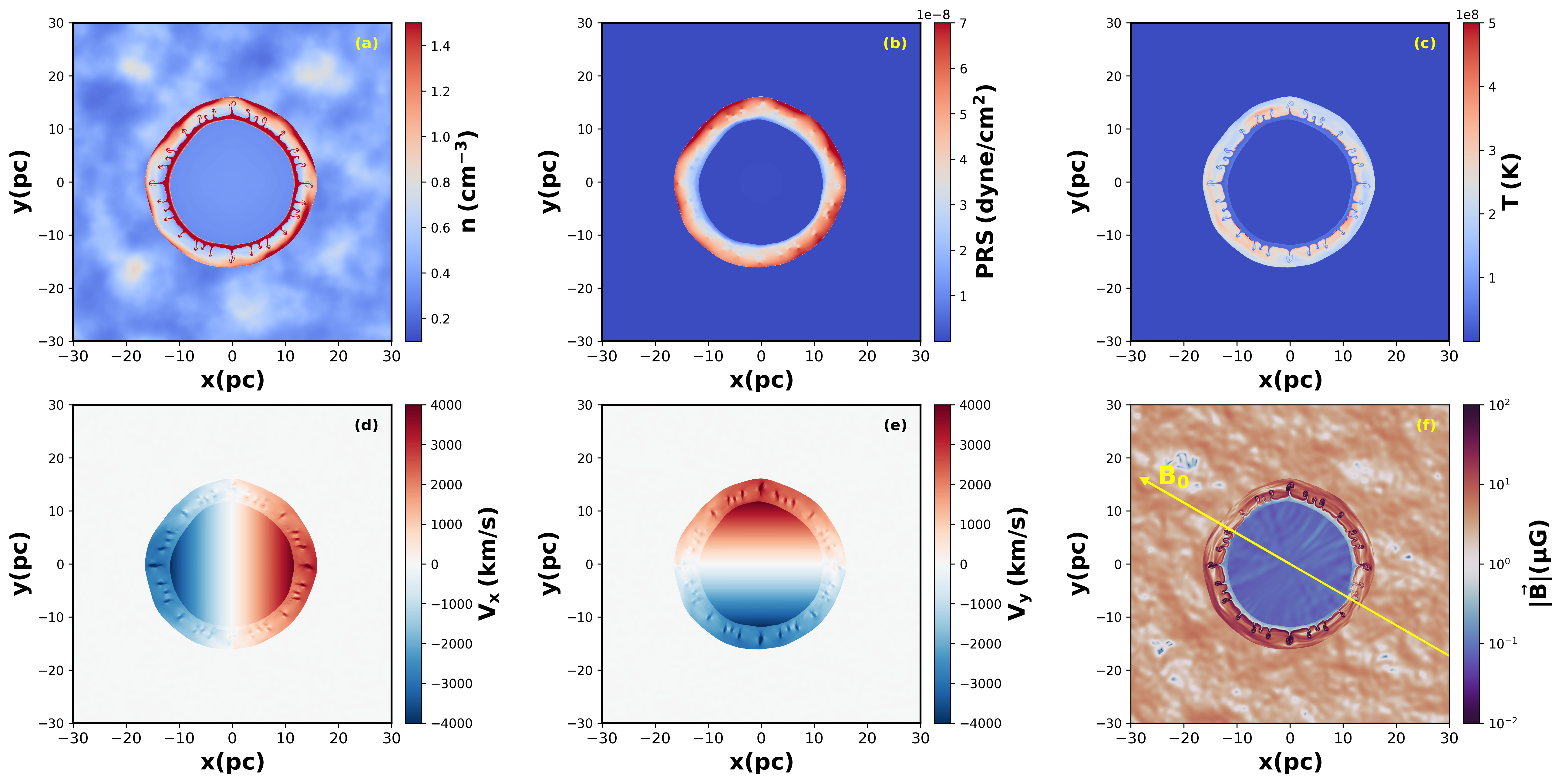}
\caption{Spatial distribution of the number density, pressure, temperature, x- and y-components of the velocity, and of the magnetic field magnitude from the 2D simulation with an age of 3000 years  for the case $M_{ej}=5 M_{\odot}$ (Run 7). }
\label{fig:maps-5Mej}
\end{figure}

We also perform simulations by varying the values of the mass of the ejecta $M_{ej}$, to study its effect on the evolution of a SNR. We make a comparison among Run 2, Run 6, and Run 7 (see Table \ref{table:table1}). In Figures \ref{fig:maps-3Mej} and \ref{fig:maps-5Mej} we report the spatial maps of the number density, pressure, temperature, the velocity components, and the magnetic field amplitude. 

At variance with the maps displayed in Figure \ref{fig:maps-sim}, it is possible to observe that the inner core is denser, especially in the case of 5$M_{ej}$ (Run 7). This means that, as the mass of the ejecta increases, the swept-up time increases. As a consequence, the radius of the FS in Run 6 and Run 7 is smaller than in Run 3. In the CD region, the expanding mass interacts with the turbulent and dense ISM environment but it also interacts with an inner dense core. This inhibits the formation of the RTI structures in that region. We further observe a lower magnetic field amplification close to the expanding shock with respect to previous simulations.
We found that the magnetic energy and the maximum value reached by the magnetic field within the simulation domain decrease as the mass of the ejecta increases. On the other hand, the kinetic energy increases slower in Run 6 and 7 and, consequently, the thermal energy decreases slower, making the expansion time longer for higher values of $M_{ej}$. Again, the level of fluctuations is high enough to distort the shock surface and create knots and filaments.

\subsection{Simulations with different background densities}
We further study the influence of the value of the mean background density on the expansion of the remnant. 

\begin{figure}[!ht]
\centering
\includegraphics[width=0.7\textwidth]{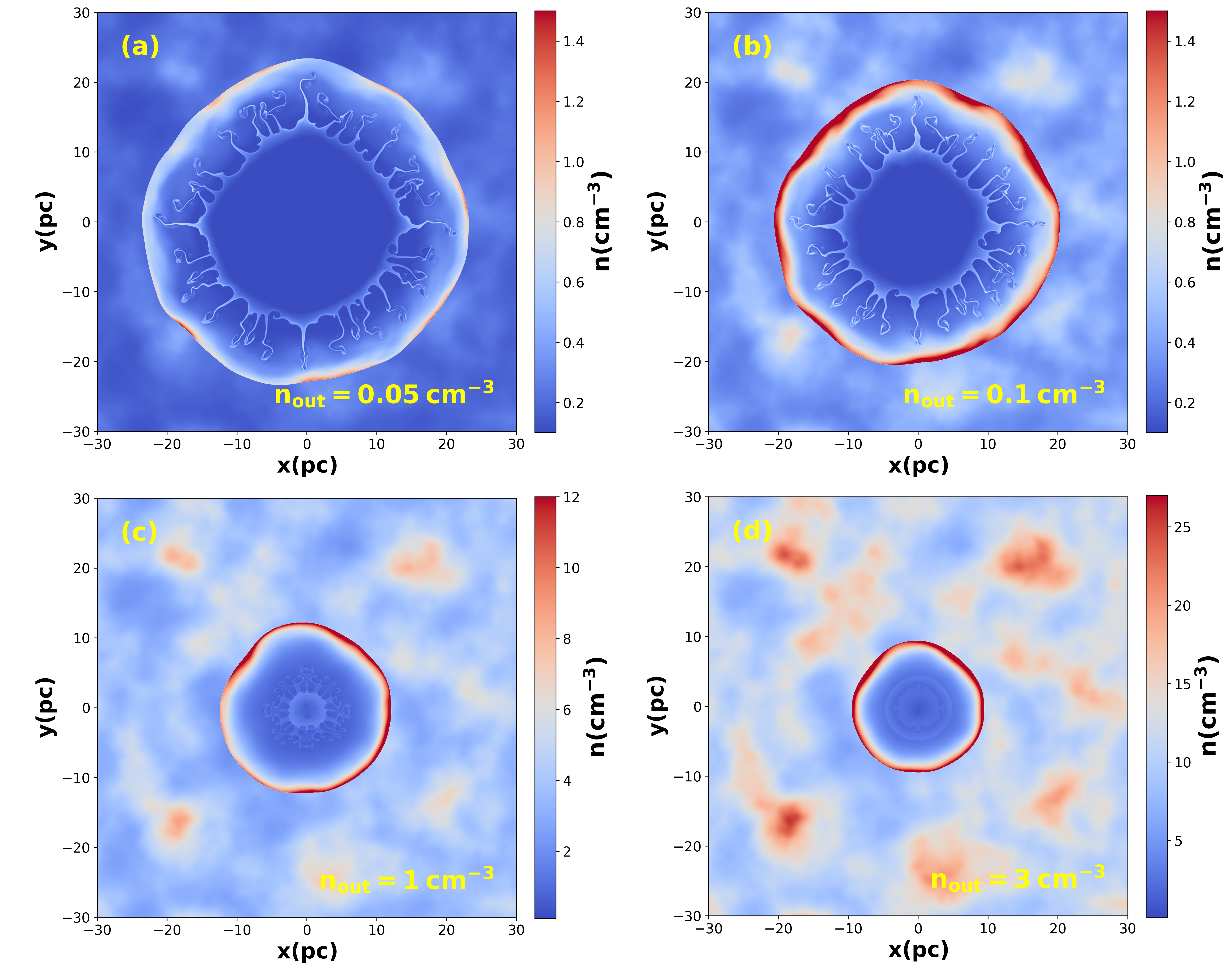}
\caption{Spatial distribution of the number density for Run 8, Run 2, Run 9 and Run 10 with an age of 3000 years. }
\label{fig:comp-nout}
\end{figure}

\begin{figure}[!h]
\centering
\includegraphics[width=0.7\textwidth]{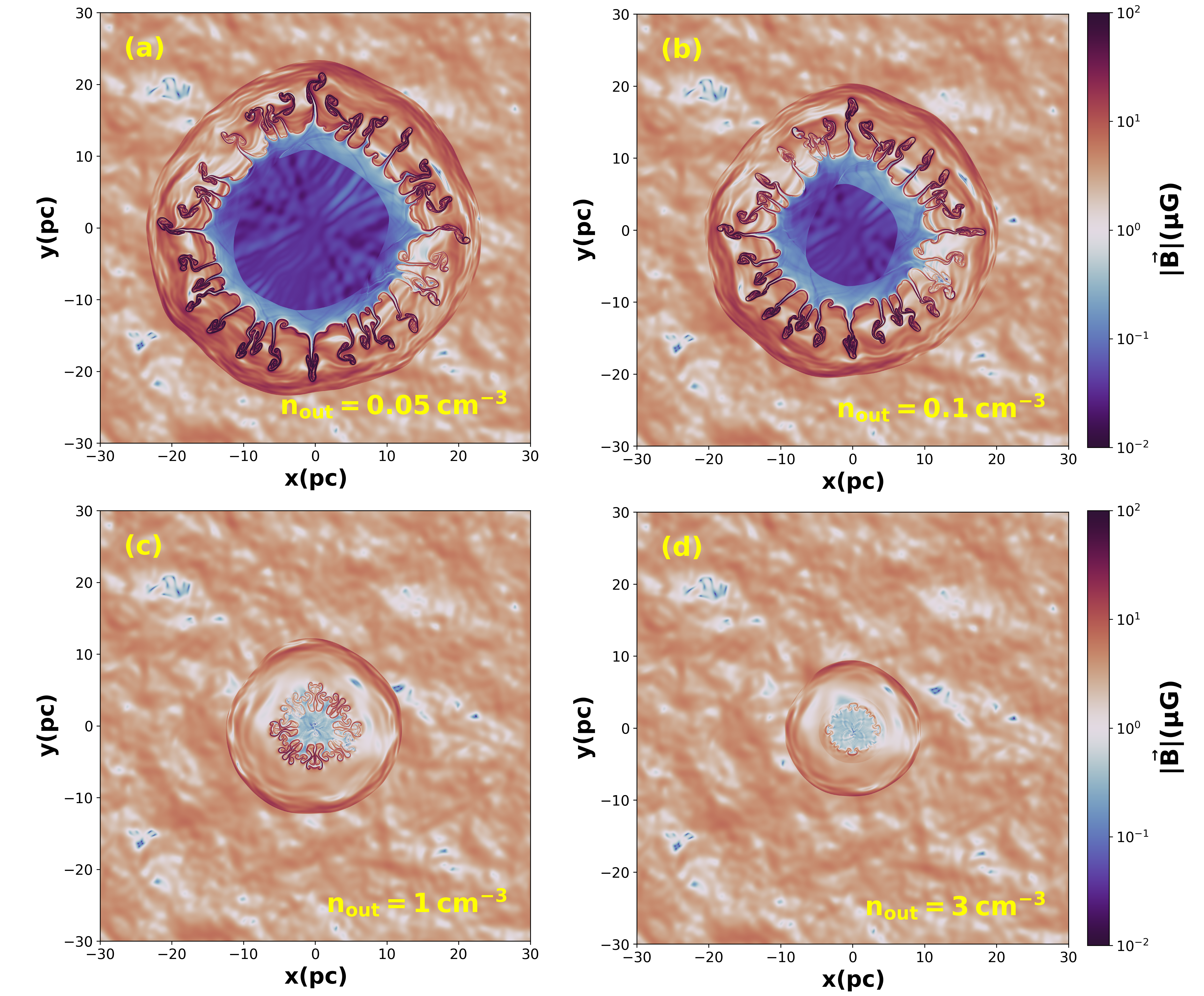}
\caption{Spatial distribution of the magnetic field magnitude for Run 8, Run 2, Run 9, and Run 10 with an age of 3000 years. }
\label{fig:comp-Bnout}
\end{figure}

When $n_{out}=0.05 $ cm$^{-3}$ (Run 8 in Table \ref{table:table1}) the SNR expands faster because it interacts with a less dense environment. As a consequence, the FS surface appears to be less distorted and more regular, with the formation of the RT instability, as shown in Figure \ref{fig:comp-nout}(a). In the case of the magnetic field map (Figure \ref{fig:comp-Bnout} (a)) the strong compression perpendicular to the mean field direction can be noticed, although the magnetic field at the FS is also amplified in small regions all around the shock surface. We underline that such a background density value is close to the one deduced from observations of SN 1006 \citep{Acero2007,Morlino2010}. A different behavior is observed in the case of higher ISM densities $n_{out}=1 $ cm$^{-3}$ and $n_{out}=3 $ cm$^{-3}$. We observe that a given age of the remnant, the radius of the SNR is smaller and the RTI has not completely developed, yet. This is due to the fact that, at the final time, the density values inside and outside the remnant are comparable, thus inhibiting the development of the RTI. 
Notice the development of magnetic field structures in the ISM that tend to be aligned with the mean magnetic field direction. \\

Also in this case we compare the averaged energies and the magnetic field strength: we find that the magnetic field strength, the magnetic energy, and the kinetic energy associated to the expansion assume higher values when the background density is lower, in agreement with the results shown in  \citet{guo2012amplification}.


\section{Comparison with Chandra observations}
A qualitative comparison between simulations and X-ray observations of SN 1006, as detected by the \textit{Chandra} spacecraft, has been performed. The X-ray data from \textit{Chandra} are retrieved from \url{https://chandra.harvard.edu/photo/openFITS/xray_data.html}. Data are provided by the Advanced CCD Imaging Spectrometer(ACIS) instrument onboard \textit{Chandra}, that gives us information about the energy, the position, and the arrival time of the X-ray photons. We used observations in the 1.34-3.0 keV energy channel. 


The X-ray emission is brighter at the edge of the supernova blast wave and it is mainly caused by non thermal synchrotron emission from relativistic electrons gyrating in an amplified magnetic field. It has been assessed that the brightest cups correspond to regions in which the shock normal direction is parallel to the direction of the magnetic field \citep{reynoso2013radio,Zhou23}.


\noindent

\begin{figure}[!ht]
\centering
\includegraphics[width=1\textwidth]{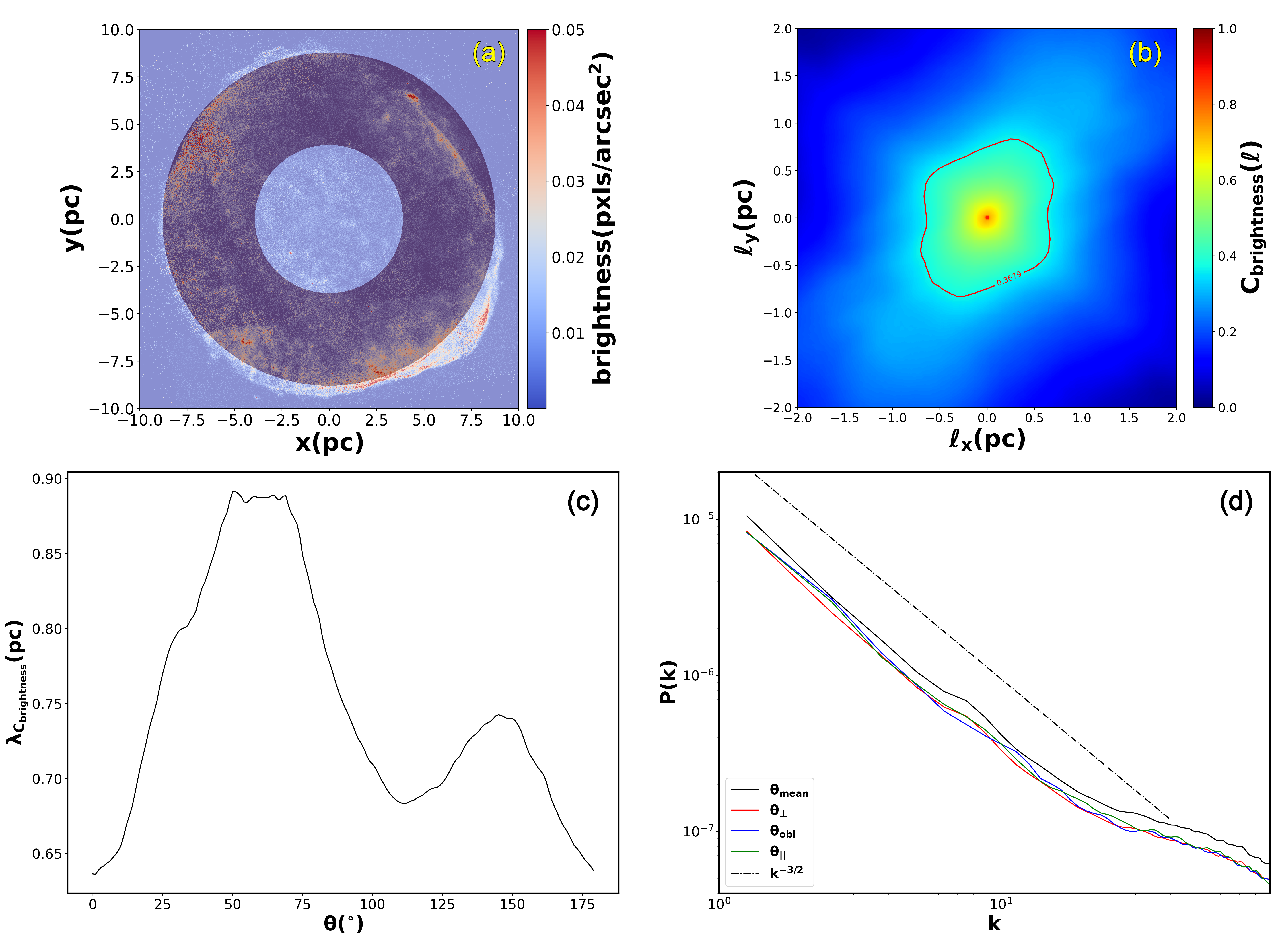}
\caption{(a) \textit{Chandra} brightness map with the ring-shaped mask overlapped. (b) Autocorrelation function map calculated on the ring-shaped region. The red solid line represents the autocorrelation length, namely the iso-contour of $C_2=1/e$. (c) Autocorrelation length as a function of angle obtained for Chandra image. (d) Spectra obtained from the analysis on the \textit{Chandra} surface brightness. }
\label{fig:chandra-mask}
\end{figure}

To make a visual comparison between Chandra data and the PLUTO results, we study the degree of anisotropy, by applying a ring-shaped mask in a similar way as described in Section \ref{sec:numerical results}, but in this case on the brightness field. The results are reported in Figure \ref{fig:chandra-mask}.

The autocorrelation function (Figure \ref{fig:chandra-mask}(b)), similar to the one reported in Figure \ref{fig:maprho}, exhibits the largest values of the correlation length along the quasi-perpendicular direction with respect to $\mathbf{B_0}$, while the smaller ones lie mainly along the quasi-parallel direction. This can also be deduced by looking at the correlation length reported in Figure \ref{fig:chandra-mask}(c). Thus, magnetic field turbulence decorrelates faster along the mean magnetic field direction. Further investigations need to be carried out in the future, since the analysis performed on the observations does not directly involve the magnetic field, as in the case of numerical simulations. 
Finally, we also determine the power spectra from the autocorrelation map. We present the three cuts in the autocorrelation map at different angles $\theta$ and the omnidirectional spectrum. In Figure \ref{fig:chandra-mask}(d) all spectra seem to have the same slope, i.e. a Kraichnan-slope with k$^{-3/2}$. Also in this case no significant spectral anisotropy has been found within the CD.

\section{Conclusions}
In this manuscript, we have tracked the evolution of a SNR using the MHD PLUTO code. Following previous works by \citet{balsara2001evolution, guo2012amplification, orlando2012role, fang2012two, yu2015three}, we run a numerical simulation in which a SNR can expand in a turbulent and dense environment. We perform a parametric study, setting up 10 types of simulations, varying the amplitude of the magnetic field fluctuations, the amplitude of the density fluctuations of the ISM, the mass of the ejecta of the SNR, and the value of the background density of the ISM. They can be considered as reference simulations also for comparison with SNR observations. 
We have analyzed in detail the simulation of a SNR with a mass of the ejecta $M_{ej}=1.4 M_{\odot}$ that evolves through an ISM with a level of magnetic fluctuations $\delta \bm{B}/\bm{B}$ = 0.5 and a level of density fluctuations of $\delta n/n=1$. The large scale magnetic field has been oriented within the $(x,y)$ plane using $\phi$=90$^{\circ}$ and $\tilde{\theta}$= 150$^{\circ}$, which is the orientation deduced from observations of SN 1006 \citep{reynoso2013radio}. 

The simulations have been run up to a final time of 3000 years, which corresponds to the Sedov-Taylor phase for a SNR. At this age, we do see the development of the RTI in the CD region. Almost 54\% of the initial thermal energy is converted into kinetic energy of the expansion. 
The presence of a highly turbulent environment favours magnetic field amplification in particular in the regions perpendicular to the mean magnetic field direction. Indeed, within a highly turbulent ISM, we observe distortion of the shock surface, resulting in the emergence of knots and filaments. We ﬁnd that, within filamentary regions close to the forward shock, the magnetic ﬁelds are of the order of 100 $\mu$G, signiﬁcantly ampliﬁed above the typical ISM value of about 3 $\mu$G. Furthermore, the presence of density fluctuations in the ISM leads to the formation of non-homogeneity in the density along the shock front. On the other hand, if the background density of the ISM is particularly high (see Run 10), the expansion of the SNR is slowed down, inhibiting the formation of RTI within the CD. 

Thus, we have investigated the effect of the turbulent environment on the level of anisotropy of the magnetic and the density fields within the SNR: this has been made by a new technique based on the application of a ring mask on the region between the RS and the FS to the density and the magnetic field maps. We determine the 2D structure function $S_2(\Bell)$ and the autocorrelation function ($C(\Bell)$) in the ring-shaped region. In order to make a more detailed statistical analysis, we interpolate in polar coordinates the autocorrelation function, from which we extrapolate the correlation length values for fixed angle $\lambda_c(\theta)$. We also determine the power spectral density at fixed angle values $P(k,\theta)$ using the Blackman-Tuckey technique.
The autocorrelation map suggests in both cases that the correlation scale of turbulence is larger in the direction quasi-perpendicular to $\mathbf{B_0}$ and shorter along it. On the other hand, for very low $\delta n/n$, isotropy is recovered in the density field within the SNR, since no density inhomogeneities are induced along the shock front. 
Then, we vary the magnetic fluctuations level and fix $\delta n/n=1$. While no significant variations are observed in the plasma quantities, the magnetic energy increases as  $\delta B/B$ increases. Thus, we obtain a noticeable amplification of the magnetic field at the shock front (mostly along the perpendicular region) for enhanced ISM turbulence. 
No significant variations are observed in the other case, in which we vary the level of the density fluctuations and we fix the level of magnetic turbulence to $\delta \bm{B}/\bm{B}=0.5$. A small difference is related to the expansion radius of the supernova which decreases as the value of $\delta n/n$ increases. Since large amplitude density fluctuations slow down the SNR expansion, the remnant is less developed and the magnetic field is less amplified. Another slightly difference is observed in the autocorrelation length values that increase as the values of $\delta n/n$ become greater.

When we run the simulations in which we varied the mass of the ejecta, we noticed that the radius of the SNR, over a time corresponding to about 3000 years, is smaller when the mass of the ejecta is higher, since it needs more time to sweep away all the mass initially concentrated in the cylindrical inner core. Further, the RTI is inhibited within the CD. 

Also, the values of the magnetic energy are smaller when the mass of the ejecta is higher. The maximum value of $\vert \bm{B} \vert$  behaves in the same manner. The kinetic energy increases at a slower rate as the mass of the ejecta increases. Indeed, when the mass of the ejecta swept up by the inner core is higher, a greater kinetic energy is required in order to wipe out all the mass. This is related to the final time that we chose for the simulations.

Variation of the background density of the ISM causes a change in the expansion rate: indeed, as the background density increases the remnant expands slower than in the simulations with low background density. Maps also show that the shock surface is highly distorted because during the expansion the remnant interacts with a very dense environment, and as a consequence, the RTI cannot fully develop and the maps show an isotropic behavior. 

Finally, we make a preliminary visual comparison between \textit{Chandra} observations of SN 1006 and numerical simulations. From the brightness map, it is clear that SN 1006 has two bright emitting regions, located on the north-eastern side and in the south-western part. These regions are dominated by synchrotron emission from relativistic electrons gyrating around the magnetic field amplified at the shock. Of course, in our PLUTO simulations, we neither take into account the cosmic ray feedback to the plasma, nor any radiation model to mimic the observed radiation at 1-3 keV. We defer the use of the particle-in-cell version of the PLUTO code \citep{mignone2018particle, vaidya2018particle} to a future work. 

We have computed the degree of anisotropy as extracted by the Chandra observations, applying a ring mask to the surface brightness. We find a qualitative good agreement between the autocorrelation maps computed over the SN 1006 brightness and the magnetic and density maps in the PLUTO simulations. The largest correlation length values are located in the region in which the FS is perpendicular to the mean magnetic field direction. Notice that this is not a one-to-one comparison because, on the one hand, the observed and simulated quantities are not the same, and, on the other hand, the inclusion of different factors, such as a modeling of the synchrotron radiation \citep{Bykov_2008,Yu2015,velazquez2017}, and the cosmic ray contribution to the energy conversion, should be taken into account in the simulations. 

We have here presented a collection of numerical simulations of an expanding SNR in a turbulent medium, by varying fundamental parameters both in the ISM and within the remnant, in order to evaluate which parameter has a major impact on the SNR evolution. Since in astrophysical systems shock waves propagate in perturbed medium \citep{Neugebauer2005,trotta2021phase, Nakanotani22}, setting up a turbulent ISM becomes a more realistic choice. 
In a future work, we plan to extend our simulations to the 3D case, since geometry does affect the magnetic field amplification in filamentary regions close to the forward shock \citep{hu2022turbulent}; we further plan to include the contribution of the cosmic rays in the magnetic field amplification and in the turbulence excitation, in order to make a better and more quantitative comparison with SN 1006 observations.

The results obtained from such simulations can help us in understanding the physical conditions in which SNRs expand in the ISM.

\textbf{Acknowledgements}
G.P. and L.P. acknowledge support by EU FP7 2007-13 through the MATERIA Project
(\texttt{PONa3\char`_00370}) and EU Horizon~2020 through the \texttt{STAR\char`_2} Project (\texttt{PON R\char`&I
2014-20, PIR01\char`_00008}) for running the simulations on the ``newton" cluster.\\
Authors acknowledge compelling discussions with Salvatore Orlando. \\
C.M. acknowledges the support from the ERC Advanced Grant ``JETSET: Launching, propagation and emission of relativistic jets from binary mergers and across mass scales'' (Grant No. 884631).\\
The simulations have been performed at the Alarico cluster, at University of Calabria, supported by “Progetto STAR 2-PIR01 00008” (Italian Ministry of University and Research). The authors acknowledge supercomputing resources and support from ICSC-Centro Nazionale di Ricerca in High Performance Computing, Big Data and Quantum Computing-and hosting entity, funded by European Union-NextGenerationEU.\\
G.P. acknowledges support by the Italian PRIN 2022, project 2022294WNB entitled "Heliospheric shocks and space weather: from multispacecraft observations to numerical modeling”.  Finanziato da Next Generation EU, fondo del Piano Nazionale di Ripresa e Resilienza (PNRR) Missione 4
“Istruzione e Ricerca” - Componente C2 Investimento
1.1, ‘Fondo per il Programma Nazionale di Ricerca e
Progetti di Rilevante Interesse Nazionale (PRIN). \\
S.P. acknowledges the project ‘Data-based predictions of solar energetic particle arrival to the Earth: ensuring space data and technology integrity from hazardous solar activity events’ (CUP H53D23011020001) ‘Finanziato dall’Unione europea – Next Generation EU’ PIANO NAZIONALE DI RIPRESA E RESILIENZA (PNRR) Missione 4 “Istruzione e Ricerca” - Componente C2 Investimento 1.1, ‘Fondo per il Programma Nazionale di Ricerca e Progetti di Rilevante Interesse Nazionale (PRIN)’ Settore PE09.\\
S. S. and S. P. acknowledge the Space It Up project funded by the Italian Space Agency, ASI, and the Ministry of University and Research, MUR, under contract n. 2024-5-E.0 - CUP n.I53D24000060005.

\bibliography{bibliography}{}
\bibliographystyle{aasjournal}



\end{document}